\documentclass[Conference,a4paper]{IEEEtran}
\IEEEoverridecommandlockouts
\usepackage{color}
\usepackage{graphicx}
\usepackage{epstopdf}
\usepackage{amsmath}
\usepackage{amssymb}
\usepackage{algorithm}
\usepackage{algorithmic}
\usepackage{amsmath}
\usepackage{multirow}
\usepackage{booktabs}
\usepackage{array}
\usepackage{amsthm}
\usepackage{stfloats}
\usepackage{caption}
\usepackage{subfigure}
\usepackage{bm}
\usepackage{booktabs}
\usepackage{setspace}
\usepackage{diagbox}
\usepackage{enumerate}
\usepackage{ulem}
\usepackage{url}
\usepackage{circledsteps}
\usepackage{pifont}
\usepackage{hyperref}
\allowdisplaybreaks[4]

{\bgroup
 \addtolength\abovedisplayshortskip{#1}
 \addtolength\abovedisplayskip{#1}
 \addtolength\belowdisplayshortskip{#1}
 \addtolength\belowdisplayskip{#1}
 }
{\egroup\ignorespacesafterend}
\newcommand{\be}{\begin{equation}}
\newcommand{\ee}{\end{equation}}
\newcommand{\bea}{\begin{eqnarray}}
\newcommand{\eea}{\end{eqnarray}}
\newcommand{\ba}{\begin{array}}
\newcommand{\ea}{\end{array}}

\newcommand{\cmark}{\ding{51}}
\newcommand{\xmark}{\ding{55}}
\title{
Fair Rate Maximization for Multi-User Multi-Cell MISO Communication Systems via
Novel Transmissive RIS Transceiver
}
\author{\IEEEauthorblockN{Yuan Guo, Wen Chen, Qingqing Wu, Zhendong Li, Kunlun Wang, Hongying Tang, and Jun Li
}
\thanks{
Y. Guo, W. Chen and
Q. Wu are with Department of Electronic Engineering, Shanghai Jiao Tong University, Shanghai, China, 
email:
yuanguo26@sjtu.edu.cn,
wenchen@sjtu.edu.cn,
qingqingwu@sjtu.edu.cn.}
\thanks{Z. Li is with the School of Information and Communication Engineering, Xi'an Jiaotong University, Xi'an, China,
email:
lizhendong@xjtu.edu.cn.}
\thanks{
K. Wang is with 
the School of Communication and Electronic Engineering, 
East China Normal University, Shanghai, China,
email: klwang@cee.ecnu.edu.cn.
}
\thanks{
H. Tang is with Science and Technology on Microsystem Laboratory, Shanghai Institute of Microsystems and Information Technology, Chinese Academy of Sciences, Shanghai, China, email: tanghy@mail.sim.ac.cn.
}
\thanks{
J. Li is with the School of Information Science and Engineering, Southeast University, Nanjing, China, 
email: jleesr80@gmail.com.
}
}

\begin{document}
\maketitle
\pagestyle{empty}
\thispagestyle{empty}

\begin{abstract}
This paper explores 
a multi-cell multiple-input single-output (MISO) downlink communication system enabled by a unique transmissive reconfigurable intelligent surface (TRIS) transceiver configuration.
Within this system framework,
we formulate an optimization problem for the purpose of maximizing the minimum rate of users for each cell via designing the transmit beamforming of the TRIS transceiver,
subject to the power constraints of each TRIS transceiver unit.
Since the objective function is non-differentiable, 
the max-min rate problem is difficult to solve.
In order to tackle this challenging optimization problem,
an efficient low-complexity optimization algorithm is developed.
Specifically,
the log-form rate function is transformed into a tractable form by employing the fractional programming (FP) methodology.
Next,
the max-min objective function can be approximated using a differentiable function derived from smooth approximation theory.
Moreover, by applying the majorization-minimization (MM) technique and examining the optimality conditions,
a solution is proposed that updates all variables analytically without relying on any numerical solvers.
Numerical results are presented to demonstrate the convergence and effectiveness of the proposed low-complexity algorithm.
Additionally, the algorithm can significantly reduce the computational complexity without performance loss.
Furthermore, the simulation results illustrate the clear superiority of the deployment of the TRIS transceiver over the benchmark schemes.

\end{abstract}


\maketitle
%\vspace{-0.25em}
\section{Introduction}

%%%（1）引入RIS
In recent years, 
the concept of reconfigurable intelligent surface (RIS) \cite{ref_RIS_1} 
has gained considerable momentum as a promising technology for sixth-generation (6G) wireless communication networks.
This technology,
which is also widely known as intelligent surface (IS) \cite{ref_RIS_3},
has attracted significant attention from both academic researchers and industry practitioners.
The unique features of RIS position it as a key enabler for overcoming fundamental challenges in future wireless systems.

%%%（2）RIS的组成架构，工作原理
Typically, 
the RIS is a planar surface formed by an extensive array of tunable elements, 
which are commonly implemented via semiconductor components such as varactor diodes and/or positive intrinsic negative (PIN) diodes.
Each unit can independently and controllably alter the phase and/or amplitude of the incident signals. 
%%%（3）RIS的优势
The inherent adaptability of RIS technology allows for its flexible integration into various intricate wireless environments, including urban canyons, indoor scenarios, and dense network deployments.
This flexibility allows RIS to effectively manipulate electromagnetic wave propagation, 
leading to significant improvements in signal coverage, quality, and overall wireless channel characteristics.
Besides,
as a passive device, 
RIS does not require active radio-frequency (RF) chains or high power consumption, 
which drastically lowers both the energy requirements and hardware complexity of the network. 
Consequently, RIS-based systems can provide a highly cost-effective and energy-efficient alternative for future wireless network deployments.

%%% RIS applications
Motivated by the numerous advantages of the RIS architecture, 
a substantial and rapidly expanding body of research has focused on investigating the deployment of RIS in wireless networks from multiple perspectives, aiming to significantly improve overall system performance, 
e.g., \cite{ref_RIS_A_1}$-$\cite{ref_RIS_A_15}.
For instance,
the authors of \cite{ref_RIS_A_1} addressed the challenge of maximizing the weighted sum-rate in RIS-enabled multi-cell systems
with the goal of enhancing downlink communication for users at the cell edge while mitigating interference across cells.
The paper \cite{ref_RIS_A_2} aimed to maximize the sum-rate of all multi-cast groups by jointly optimizing base station (BS) precoding and RIS reflection coefficients.
Two efficient algorithms with second-order cone program (SOCP) and closed-form solutions are proposed, 
and numerical results demonstrate significant improvements in spectral and energy efficiency with reduced computational complexity.
The work \cite{ref_RIS_A_3}  adopted the RIS to enhance the sum-rate of the multi-cell non-orthogonal multiple access (NOMA) networks by jointly optimizing user association, resource allocation, and RIS phase shifts, 
achieving significant improvements in sum-rate and energy efficiency.
The literature \cite{ref_RIS_A_4} designed joint transmit and reflective beamforming for RIS-aided multi-cell multiple-input single-output (MISO) systems using an alternating optimization algorithm. 
It outperforms the benchmark zero-forcing scheme and ensures user fairness via signal-to-interference-plus-noise ratio (SINR) balancing.
The authors in \cite{ref_RIS_A_5} studied an RIS-assisted secure multi-user communication system with hardware impairments,
aiming to maximize the weighted minimum approximate ergodic secrecy rate.
They proposed both SOCP-based and low-complexity algorithms to efficiently solve this problem.
In \cite{ref_RIS_A_6},
a secure RIS-assisted simultaneous wireless information and power transfer (SWIPT) network with arbitrary information and energy receivers was studied,
where the weighted sum transferred power is maximized via a novel iterative algorithm.
The work \cite{ref_RIS_A_6_1} investigated a novel double-faced active (DFA)-RIS structure-aided secure communication system.
The paper \cite{ref_RIS_A_6_2} considered an elements allocation problem in a joint active and passive RIS-assisted communication system.
A dual-RIS aided SWIPT system was researched in \cite{ref_RIS_A_6_3}.
Based on statistical channel state information (CSI),
a low-complexity phase-shift optimization and power allocation method for RIS-aided multi-cell massive 
multiple-input multiple-output (MIMO) systems was proposed in \cite{ref_RIS_A_7},
providing closed-form rate formulas and user-fairness guarantees.
The authors of \cite{ref_RIS_A_8} introduced an end-to-end deep learning beamforming approach
for RIS-assisted wideband MIMO systems
that operates without explicit CSI,
and proposed both true time delay (TTD)-based and subarray-based RIS architectures to mitigate near-field beam split and improve spectral efficiency.
The paper \cite{ref_RIS_A_9} demonstrated
that a novel distortion-and-aging-aware minimum-mean-square-error (DAA-MMSE) receiver proposed for an RIS-aided multi-cell massive MIMO system significantly increases spectral efficiency while reducing the pilot overhead.
The work \cite{ref_RIS_A_10} developed a low-complexity solution for RIS-aided full-duplex integrated sensing and communication (ISAC) systems, 
addressing joint optimization challenges and demonstrating notable performance gains via RIS deployment. 
The literature \cite{ref_RIS_A_11} integrated the novel intelligent omni surface (IOS) architecture into the ISAC system to achieve the full-view coverage,
aiming to maximize the minimum sensing SINR while guaranteeing the quality of multi-user communications.
The authors in \cite{ref_RIS_A_12} considered the sum-rate maximization problem
in multi-cell systems assisted by simultaneously transmitting and reflecting (STAR)-RIS with multiple operational modes.
The beyond diagonal (BD)-RIS assisted multi-band multi-cell MIMO system considering the frequency-dependent properties was researched in \cite{ref_RIS_A_13},
which demonstrates the superior performance of BD-RIS over traditional single-connected designs.
The authors of \cite{ref_RIS_A_14} 
introduced an RIS into a cooperative multi-cell ISAC network containing multi-user and multi-target
to enhance communication and sensing performances.
The paper \cite{ref_RIS_A_15} studied RIS-assisted multi-cell MIMO networks that combine over-the-air (OTA) computation 
and aim at minimizing the MSE.

Beyond the typical RIS employed as an auxiliary module in wireless systems,
a novel transmissive RIS (TRIS) transceiver architecture was presented in \cite{ref_TRIS_1}.
The proposed TRIS transceiver architecture differs from conventional multi-antenna transmitters 
by integrating a passive transmissive RIS alongside a single horn antenna feed, 
thereby avoiding the use of numerous RF chains and complex signal processing units, 
and achieving superior system performance with lower power consumption.
In addition,
relative to the reflective RIS transmitters detailed in \cite{ref_RRIS_1}$-$\cite{ref_RRIS_2},
the unique TRIS transceiver design is capable of resolving the following two major problems:

\noindent
\textit{1) feed source blockage}:
For a reflective-type RIS transmitter, 
positioning both the horn antenna and the user on the same side of the RIS 
causes a feed source blockage effect on the incident electromagnetic (EM) wave.
In the TRIS transceiver architecture,
the horn antenna is placed on one side of the RIS, 
while the user is located on the opposite side.
Therefore, this effect can be eliminated;

\noindent
\textit{2) echo interference}:
Reflective RIS transceivers suffer from echo interference since the incident and reflected EM waves coexist on the same side of the RIS. 
The TRIS transceiver architecture alleviates this challenge 
by spatially separating the incident and transmitted waves across opposite sides of the RIS.
Therefore, 
the TRIS transceiver represents a novel technology facilitating sustainable capacity growth with improved cost efficiency.

\subsection{Prior Works}

Leveraging the advantages of the TRIS transceiver architecture, 
recent works have explored TRIS transceiver-assisted wireless networks 
across different aspects to boost overall system performance, 
e.g., \cite{ref_TRIS_A_1}$-$\cite{ref_TRIS_A_9}.
For instance,
the authors of \cite{ref_TRIS_A_1} investigated
a TRIS transceiver-enhanced multi-stream downlink communication framework 
leveraging the time-modulated array (TMA) technique, 
and proposed a linear-complexity algorithm to solve the max-min SINR optimization problem.
The paper \cite{ref_TRIS_A_2} adopted a TRIS transceiver-based receiver architecture for the uplink communication system, in which uplink users employ orthogonal frequency division multiple access (OFDMA).
Besides,
\cite{ref_TRIS_A_2} aimed to maximize the sum-rate of uplink users while guaranteeing the individual quality-of-service (QoS).
Focusing on the TRIS transceiver-assisted SWIPT system,
the work \cite{ref_TRIS_A_3} studied the sum-rate maximization problem 
and presented an algorithm whose enhanced performance was demonstrated through simulation results.
The literature \cite{ref_TRIS_A_4} 
aimed to minimize the total energy consumption while satisfying communication and computing resource requirements
in a TRIS transceiver-aided multi-tier computing network.
The authors in \cite{ref_TRIS_A_5} designed a hybrid TRIS transceiver framework combining active and passive RIS elements, 
each capable of switching modes dynamically. 
Their numerical analysis revealed notable improvements in energy efficiency (EE) for downlink multi-user communication scenarios.
In \cite{ref_TRIS_A_6}, 
the authors utilized the TRIS transceiver to enable multi-beam transmission alongside directional beam suppression 
by optimizing a max-min objective subject to nonlinear constraints.
Furthermore, 
to connect the beamforming approach with practical implementation, 
the study proposed a realistic model capturing the TRIS transceiver's behavior at both input and output.
A TRIS transceiver-aided secure communication system was researched in \cite{ref_TRIS_A_6_1},
which validated that the TRIS transceiver can significantly improve the weighted sum secrecy rate.
The authors of \cite{ref_TRIS_A_7} developed a time-division sensing and communication protocol for a TRIS transceiver-assisted robust and secure ISAC system. 
In addition, rate-splitting multiple access (RSMA) was employed to enhance interference management and bolster security against eavesdropping.
The paper \cite{ref_TRIS_A_8} investigated a distributed cooperative ISAC network aided by the TRIS transceiver to improve service coverage. 
The primary objective of the study was to maximize the minimum radar mutual information (RMI) to enhance system performance.
The work \cite{ref_TRIS_A_9} investigated the maximization of sum-rate performance for multi-cluster communications in a Low Earth Orbit (LEO) satellite system employing nonorthogonal multiple access (NOMA) and leveraging the TRIS transceiver architecture.
{
The comparison between this work and existing works is summarized in Table \ref{Comparison of Existing Works}.
}

\begin{table*}[t]
%\begin{small}
\centering
\caption{{Comparison of Existing Works}}
{
\begin{tabular}{|c|c|c|c|} \hline \label{Comparison of Existing Works}
References                & Beamforming Design & Low-Complexity Method  & Multi-Cell System      \\ \hline
\cite{ref_TRIS_A_2}, \cite{ref_TRIS_A_3}, \cite{ref_TRIS_A_4}, \cite{ref_TRIS_A_5}, \cite{ref_TRIS_A_6_1}, \cite{ref_TRIS_A_7}, \cite{ref_TRIS_A_9} & \cmark             & \xmark                    & \xmark                            \\ \hline
\cite{ref_TRIS_A_1}, \cite{ref_TRIS_A_6}, \cite{ref_TRIS_A_8} & \cmark             & \cmark                   & \xmark                            \\ \hline
This paper          & \cmark             & \cmark                        & \cmark               \\ \hline
\end{tabular}}

%\end{small}
\end{table*}

\subsection{Motivations and Contributions}

Nevertheless, 
it is obvious that most of the works mentioned above related to
the TRIS transceiver \cite{ref_TRIS_A_1}$-$\cite{ref_TRIS_A_9} 
have focused on single-cell scenarios, 
whereas the general scenario involving multi-cell configurations remains unexplored.
It is well known that inter-cell interference is a non-negligible issue for multi-cell users, leading to degraded system performance.

In this work,
we deploy the TRIS transceiver architecture in the multi-cell communication system with multiple users, as shown in Fig. \ref{fig.1}.
Specifically,  
the contributions of this paper are summarized as follows:
\begin{itemize}
\item
This paper studies the beamforming design in a multi-cell MISO downlink communication system enhanced by the advanced TRIS transceiver architecture 
to boost wireless communication performance and system reliability significantly.
Specifically,
the objective is to maximize the minimum rate of each cell by designing TRIS transceiver beamforming,
constrained by the per-element maximum transmit power of the TRIS transceiver.
To the best of our knowledge, 
this problem has not yet been investigated in the existing literature,
e.g., \cite{ref_TRIS_A_1}$-$\cite{ref_TRIS_A_9}.

\item
{
Due to the complex and non-differentiable nature of the objective function, 
solving the highly non-convex max-min rate problem is particularly challenging. 
In order to effectively tackle this optimization problem,
by using the fractional programming (FP) framework,
we first convert the original rate function into a more tractable form.
And then,
we approximate its max-min objective function with a differentiable approximation function based on smooth approximation theory.
Furthermore,
by leveraging the majorization-minimization (MM) method and analyzing optimality conditions,
we successfully develop a low-complexity algorithm that updates all variables analytically and does not rely on any numerical solvers.
}

\item
Last but not least,
extensive simulation results are provided to demonstrate the effectiveness and efficiency of our proposed low-complexity algorithm,
which can greatly lower computational complexity without sacrificing communication performance.
Furthermore, 
the results validate that the deployment of the TRIS transceiver markedly improves the performance of multi-cell communication systems.
In addition, 
when compared with the conventional transceiver, 
the TRIS transceiver using the proposed algorithm requires only approximately 66\% of the power consumption to achieve a similar sum-rate level.

\end{itemize}

The rest of the paper is organized as follows. 
Section II will introduce the  model of the TRIS transceiver-enabled multi-cell communication system
and propose the max-min rate optimization problem.
Section III will propose a low-complexity solution to tackle the proposed problem.
Sections IV and V will present numerical results and conclusions of the paper, respectively.

%\textit{Notations:}
\subsection{Notations}
Lower-case and boldface capital letters are represented as 
vectors and matrices;
$\mathbf{X}^{\ast}$,
$\mathbf{X}^{T}$,
and
$\mathbf{X}^{H}$
denote the conjugate, transpose, and
conjugate transpose of matrix $\mathbf{X}$, respectively;
%%%%%
$\mathbb{C}^{N \times 1}$ represents the set of $N \times 1$ complex vectors;
%%%%%
$\mathbf{0} $ denotes the all zeros matrix;
%%%%%
$\Vert \mathbf{x} \Vert_{2}$ denotes the $l_2$ norm of the vector $\mathbf{x}$;
%%%%%
%%%%%
$ \triangleq $ and $ \sim $ signify ``defined as''
and ``distributed as'', respectively;
%%%%%
%%%%%
$ \mathbb{E}[\cdot]$ denotes the statistical expectation;
%%%%%
$\mathcal{CN}(\mathbf{x}, \boldsymbol{\Sigma})$
denotes the distribution of a circularly symmetric complex Gaussian (CSCG)
vector with mean vector $\mathbf{x}$
and covariance matrix $\boldsymbol{\Sigma}$;
%%%%%
%%%%%
$\text{diag}(\mathbf{x})$ denotes a diagonal matrix whose 
diagonal entries are given by the elements of the vector $\mathbf{x}$;
%%%%%
$\text{blkdiag}(\mathbf{X}_1, \cdots, \mathbf{X}_N)$
represents a block diagonal matrix 
with $\mathbf{X}_1, \cdots, \mathbf{X}_N$ as its diagonal blocks.

\section{System Model and Problem Formulation}
\subsection{System Model}

\begin{figure}[t]
	\centering
	\includegraphics[width=.48\textwidth]{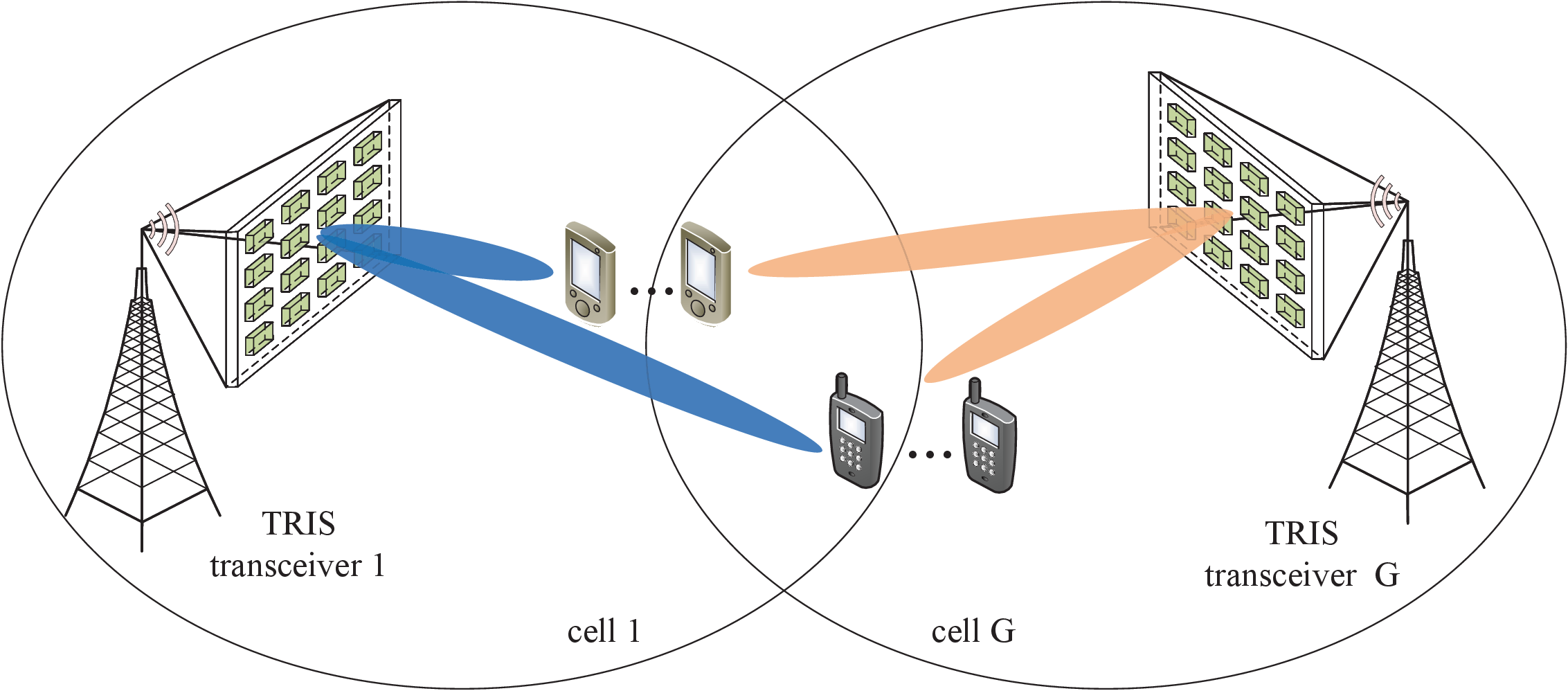}
	\caption{{An illustration of the TRIS transceiver enabled multi-cell MISO communication system.}}
	\label{fig.1}
\end{figure}

As shown in Fig. \ref{fig.1},
we consider a TRIS transceiver-enabled downlink multi-user multi-cell communication system consisting of $G$ cells.
Each cell has a TRIS transceiver equipped with $N$ units and $K$ single-antenna downlink mobile users.
Let 
$\mathcal{G} \triangleq \{1,\cdots,G\}  $,
$\mathcal{K} \triangleq \{1,\cdots,K\}  $
and
$\mathcal{N} \triangleq \{1,\cdots,N\}  $
denote the set of cells,
each cell's users
and TRIS transceiver units, 
respectively.

The transmit signal 
\footnote{
{
The signal generation mechanism of the TRIS transceiver can be found in \cite{ref_TRIS_1}.
}
}
of the $g$-th TRIS transceiver is written as
\begin{align}
\mathbf{x}_g
={\sum}_{k=1}^{K}\mathbf{f}_{g,k}s_{g,k},
\forall k \in \mathcal{K},
\forall g \in \mathcal{G},
\end{align}
where
$s_{g,k}$  denotes the data symbol of the $k$-th user in the $g$-th cell
and satisfies $\mathbb{E}[\vert s_{g,k}\vert] = 1$
and
$\mathbb{E}[ s_{g,k} s_{i,j}^{\ast}] = 0$,
$\{g,k\} \neq \{i,j\}$,
and
the vector
$\mathbf{f}_{g,k} \in \mathbb{C}^{N \times 1}$ represents the beamformer of the $g$-th TRIS transceiver for transmitting $s_{g,k}$. 

{
Furthermore,
based on the nature of the TRIS transceiver \cite{ref_TRIS_1},
the following each TRIS transceiver unit linear power constraint 
\footnote{
{
The non-linear per-unit power constraint \cite{ref_Non-linear_1}$-$\cite{ref_Non-linear_2} will be investigated in the future.
}
}
will be held for
the beamforming vectors,
which is given as}
\begin{align}
\mathbf{f}_{g}^{H}\mathbf{\bar{A}}_{n}\mathbf{f}_{g} \leq P_t,
\forall n \in \mathcal{N},
\forall g \in \mathcal{G},
\end{align}
where
$\mathbf{f}_{g} \triangleq [\mathbf{f}_{g,1}^{T}, \mathbf{f}_{g,2}^{T}, \cdots,
\mathbf{f}_{g,K}^{T} ]^T \in \mathbb{C}^{N \cdot K \times 1}$,
$P_t$ represents the TRIS transceiver unit's maximum achievable power.
A selection matrix can be given as 
\begin{align}
 \mathbf{\bar{A}}_n \triangleq \text{blkdiag}(  \mathbf{A}_n,\cdots ,\mathbf{A}_n  )\in \mathbb{R}^{N\cdot K\times N\cdot K}.
\end{align}
The diagonal submatrix
$\mathbf{A}_n \triangleq \textrm{diag}(\mathbf{a}_n )\in \mathbb{R}^{N\times N}$
is defined using the index vector $\mathbf{a}_n$,
which can be expressed as
\begin{align}
\mathbf{a}_n \triangleq [0,0,\cdots,\underbrace{1}\limits_{\textrm{n-}th},\cdots,0]^T \in \mathbb{R}^{N \times 1},
\end{align}
where 
the entry at the $n$-th  position is $1$ and all other entries are $0$.

{
The equivalent baseband channel of the TRIS transceiver of the $i$-th cell to the $k$-th user belonging to cell $g$
is denoted by
$\mathbf{h}_{i,g,k}  \in \mathbb{C}^{N\times 1} $
and is assumed to follow the quasi-static flat-fading model.
By adopting the Rician channel model \footnote{ {  
Here, we assume an ideal spatially uncorrelated channel, 
while the more practical case  of spatially correlated channels \cite{ref_spatially correlated} is left for future work.
}},
the channel $\mathbf{h}_{i,g,k}$  can be formulated as
\begin{align}
\mathbf{h}_{i,g,k}
=
\sqrt{C_0 \bigg (\frac{d}{d_0} \bigg)^{-\alpha}}
\bigg(
\sqrt{\frac{\kappa}{\kappa+1}}\mathbf{h}_{i,g,k}^{LoS}
+
\sqrt{\frac{1}{\kappa+1}}\mathbf{h}_{i,g,k}^{NLoS}
\bigg),
\end{align}
where
$C_0$ is the large-scale fading coefficient with reference distance $d_0 = 1$m.  
$d$ and $\alpha$ denote the distance and the large-scale fading factor of the corresponding channel, 
respectively.
$\kappa$ is the Rician factor.
The terms $\mathbf{h}_{i,g,k}^{LoS}$ and $\mathbf{h}_{i,g,k}^{NLoS}$ are the line-of-sight (LoS) and non-line-of-sight (NLoS) components, respectively.}
{
The above channel can be obtained by channel estimation \cite{ref_Channel_estimation}.
}

Next,
the received signal at the $k$-th user belonging to cell $g$ is formulated as
\begin{align}
y_{g,k} &= 
{\sum}_{i=1}^{G}
\mathbf{{h}}_{i,g,k}^H\mathbf{x}_i + n_{g,k}\\
&= {\sum}_{i=1}^{G}
\mathbf{{h}}_{i,g,k}^H
\bigg( {\sum}_{k=1}^{K}\mathbf{f}_{i,k}s_{i,k} \bigg)
+ n_{g,k}\nonumber\\
&= \underbrace{\mathbf{{h}}_{g,g,k}^H\mathbf{f}_{g,k}s_{g,k}}
\limits_{\textrm{Desired signal}}
+ 
\underbrace{
{\sum}_{j\neq k}^{K} \mathbf{{h}}_{g,g,k}^H\mathbf{f}_{g,j}s_{g,j}
}
\limits_{\textrm{Intra-cell interference}}
\nonumber \\
&+ 
\underbrace{
{\sum}_{i\neq g}^{G}{\sum}_{j=1}^{K}
\mathbf{{h}}_{i,g,j}^H\mathbf{f}_{i,j}s_{i,j}
}
\limits_{\textrm{Inter-cell interference}}
+ n_{g,k},\nonumber
\end{align}
where
$n_{g,k} \sim \mathcal{CN}(0,\sigma_{g,k}^2)$
represents
the complex additive white Gaussian noise (AWGN) for user $k$
in the $g$-th cell.
Let
$\mathbf{\bar{h}}_{i,j,k} \triangleq [\mathbf{{h}}_{i,j,k}^T, \cdots,\mathbf{{h}}_{i,j,k}^T]^T\in\mathbb{C}^{N\cdot K\times 1} $.
The received signal $y_{g,k}$ can be rewritten by
\begin{align}
y_{g,k} &= 
\underbrace{\mathbf{\bar{h}}_{g,g,k}^H  \mathbf{B}_{k}  \mathbf{f}_{g}s_{g,k}}
\limits_{\textrm{Desired signal}}
+ 
\underbrace{
{\sum}_{j\neq k}^{K} \mathbf{\bar{h}}_{g,g,k}^H \mathbf{B}_{j} \mathbf{f}_{g}s_{g,j}
}
\limits_{\textrm{Intra-cell interference}} \\
&+ 
\underbrace{
{\sum}_{i\neq g}^{G}{\sum}_{j=1}^{K}
\mathbf{\bar{h}}_{i,g,j}^H \mathbf{B}_{j} \mathbf{f}_{i}s_{i,j}
}
\limits_{\textrm{Inter-cell interference}}
+ n_{g,k},\nonumber
\end{align}
where
$\mathbf{B}_k \triangleq \text{diag}( \mathbf{b}_k ) \in \mathbb{R}^{N\cdot K\times N\cdot K}$ is a selection matrix,
and
$\mathbf{b}_k \in \mathbb{R}^{N\cdot K\times 1}$
is a vector defined as
\begin{align}
\mathbf{b}_k \triangleq [0,\cdots ,0,\underbrace{1,\cdots,1}\limits_{\textrm{N}},0,\cdots,0],
\end{align}
meaning that the entries from positions 
$((k-1)\times N +1 )\sim (k\times N)$ are set to 1, while all other entries are 0. 

Then,
the SINR of the $k$-th user in cell $g$ is obtained as
\begin{align}
&\text{SINR}_{g,k}\\
&=
\frac{ \vert \mathbf{\bar{h}}_{g,g,k}^H  \mathbf{B}_{k}  \mathbf{f}_{g} \vert^2 }
{{\sum}_{j\neq k}^{K} \vert\mathbf{\bar{h}}_{g,g,k}^H \mathbf{B}_{j} \mathbf{f}_{g}\vert^2
+
{\sum}_{i\neq g}^{G}{\sum}_{j=1}^{K}
\vert
\mathbf{\bar{h}}_{i,g,j}^H \mathbf{B}_{j} \mathbf{f}_{i}
\vert^2
+ \sigma_{g,k}^2
  },\nonumber
\end{align}
and the achievable rate of each user is given by
\begin{align}
\mathrm{R}_{g,k}(\{\mathbf{f}_{g}\}) = \textrm{log}(1+\text{SINR}_{g,k}), 
\forall k \in \mathcal{K}, 
\forall g \in \mathcal{G}.
\end{align}

\subsection{Problem Formulation}

To enhance rate fairness in the multi-cell MISO system,
we consider the max-min fairness problem with the goal of maximizing the minimum rate of all users in each cell
via optimizing the transmit beamformer vectors $\{\mathbf{f}_g\}$,
subject to the individual transmit power constraints at the TRIS transceiver units.
Therefore, 
the min-weighted-rate maximization problem can be formulated as
\begin{subequations}
\begin{align}
\textrm{(P0)}:&\mathop{\textrm{max}}
\limits_{\{\mathbf{f}_g\}
}\
\bigg\{ \mathrm{R}_{s} (\{\mathbf{f}_g\}) 
= 
{\sum}_{g=1}^{G} 
\mathop{\textrm{min}}
\limits_{ k \in \mathcal{K} } 
\mathrm{R}_{g,k}(\{\mathbf{f}_g\}) 
 \bigg\}\label{P0_obj}\\
\textrm{s.t.}\ 
& \mathbf{f}_{g}^H\mathbf{\bar{A}}_n\mathbf{f}_{g} \leq P_t, \forall n \in \mathcal{N}, \forall g \in \mathcal{G}.\label{P0_c_1}
\end{align}
\end{subequations}

The problem (P0) is highly challenging to tackle
due to its highly non-differentiable and non-convex objective function.

%\vspace{-0.2cm}
\section{Low-Complexity Algorithm}

\subsection{Problem Reformulation}

To make the problem (P0) more tractable,
we will adopt the fractional programming (FP) framework \cite{ref_FP}$-$\cite{ref_FP_2}
to equivalently convert the objective function (\ref{P0_obj}).
First,
by applying the Lagrangian dual reformulation and introducing auxiliary variables
$\{\gamma_{g,k}\}$,
the original rate function $\mathrm{R}_{g,k}(\{\mathbf{f}_g\}) $
can be transformed into (\ref{FP_1}).
\begin{figure*}
%\begin{small}
\begin{align}
\mathrm{\dot{R}}_{g,k}(\{\mathbf{f}_g\}, \gamma_{g,k})
=
\text{log}(1+\gamma_{g,k}) - \gamma_{g,k}
+
\frac{(1+\gamma_{g,k}) \vert \mathbf{\bar{h}}_{g,g,k}^H  \mathbf{B}_{k}  \mathbf{f}_{g} \vert^2}
{{\sum}_{i=1}^{G}{\sum}_{j=1}^{K}
\vert
\mathbf{\bar{h}}_{i,g,j}^H \mathbf{B}_{j} \mathbf{f}_{i}
\vert^2
+ \sigma_{g,k}^2
},\label{FP_1}
\end{align}
%\end{small}
\boldsymbol{\hrule}
\end{figure*}
Furthermore,
by leveraging the quadratic transform with 
introducing the auxiliary variables 
$\{\omega_{g,k}\}$,
the function
$\mathrm{\dot{R}}_{g,k}(\{\mathbf{f}_g\}, \gamma_{g,k})$
can be further rewritten in (\ref{FP_2}).
\begin{figure*}
%\begin{small}
\begin{align}
&\mathrm{\ddot{R}}_{g,k}(\{\mathbf{f}_g\}, \gamma_{g,k}, \omega_{g,k})
=
\text{log}(1+\gamma_{g,k}) - \gamma_{g,k}\label{FP_2}\\
&+2\sqrt{(1+\gamma_{g,k} )}
\text{Re}\{\omega_{g,k}^{\ast} \mathbf{\bar{h}}_{g,g,k}^H  \mathbf{B}_{k}  \mathbf{f}_{g}\}
-
\vert \omega_{g,k} \vert^2
\bigg( {\sum}_{i=1}^{G}{\sum}_{j=1}^{K}
\vert
\mathbf{\bar{h}}_{i,g,j}^H \mathbf{B}_{j} \mathbf{f}_{i}
\vert^2
+ \sigma_{g,k}^2  \bigg)
\nonumber
\end{align}
%\end{small}
\boldsymbol{\hrule}
\end{figure*}
Based on the above transformation,
the original optimization problem (P0) can be equivalently reformulated as
\begin{subequations}
\begin{align}
\textrm{(P1)}:&\mathop{\textrm{max}}
\limits_{\{\mathbf{f}_g\},
\{\gamma_{g,k}\}, 
\{\omega_{g,k}\}
}\
\bigg\{ \mathrm{R}_{s} 
= 
{\sum}_{g=1}^{G} 
\mathop{\textrm{min}}
\limits_{ k \in \mathcal{K} } 
\mathrm{\ddot{R}}_{g,k}
 \bigg\}\label{P1_obj}\\
\textrm{s.t.}\ 
& \mathbf{f}_{g}^H\mathbf{\bar{A}}_n\mathbf{f}_{g} \leq P_t, \forall n \in \mathcal{N}, \forall g \in \mathcal{G}.\label{P1_c_1}
\end{align}
\end{subequations}

In the following,
we will develop an algorithm based on the block coordinate ascent (BCA) \cite{ref_BCA}$-$\cite{ref_BCA_1} framework to solve the problem (P1).

\subsection{Optimizing auxiliary variables}

Following the derivation of the FP method \cite{ref_FP},
when other variables are given,
we can obtain the closed solutions of the auxiliary variables 
$\{\gamma_{g,k}\}$
and
$\{\omega_{g,k}\}$,
which are respectively given as
\begin{align}
&\gamma_{g,k}^{\star}
\!=\!
\frac{\vert \mathbf{\bar{h}}_{g,g,k}^H  \mathbf{B}_{k}  \mathbf{f}_{g} \vert^2}
{{\sum}_{j\neq k}^{K} \vert\mathbf{\bar{h}}_{g,g,k}^H \mathbf{B}_{j} \mathbf{f}_{g}\vert^2
\!+\!
{\sum}_{i\neq g}^{G}{\sum}_{j=1}^{K}
\vert
\mathbf{\bar{h}}_{i,g,j}^H \mathbf{B}_{j} \mathbf{f}_{i}
\vert^2
\!+\! \sigma_{g,k}^2
}, \label{FP_auxiliary_solution_gamma} \\
&
\omega_{g,k}^{\star}
=
\frac{\sqrt{(1+\gamma_{g,k} )} \mathbf{\bar{h}}_{g,g,k}^H  \mathbf{B}_{k}  \mathbf{f}_{g}   }
{{\sum}_{i=1}^{G}{\sum}_{j=1}^{K}
\vert
\mathbf{\bar{h}}_{i,g,j}^H \mathbf{B}_{j} \mathbf{f}_{i}
\vert^2
+ \sigma_{g,k}^2}.\label{FP_auxiliary_solution_omega}
\end{align}

\subsection{Updating The Beamformer}
In this subsection, 
we investigate the optimization of the beamformer $\{\mathbf{f}_g\}$ 
when \textit{the auxiliary variables} are given. 
By defining the new coefficients as follows
\begin{align}
&c_{1,g,k}
\triangleq
\text{log}(1+\gamma_{g,k}) - \gamma_{g,k}
-
\vert \omega_{g,k} \vert^2\sigma_{g,k}^2,\\
&\mathbf{b}_{1,g,k}
\triangleq
\sqrt{1+\gamma_{g,k}}\omega_{g,k}\mathbf{B}_{k}^H\mathbf{\bar{h}}_{g,g,k},\nonumber\\
&
\mathbf{B}_{1,i,g,k}
\triangleq
\vert \omega_{g,k} \vert^2
\bigg({\sum}_{j=1}^{K}
\mathbf{B}_{j}^H
\mathbf{\bar{h}}_{i,g,j}
\mathbf{\bar{h}}_{i,g,j}^H 
\mathbf{B}_{j}
 \bigg),\nonumber
\end{align}
the function 
$\mathrm{\ddot{R}}_{g,k}(\{\mathbf{f}_g\}, \gamma_{g,k}, \omega_{g,k})$
can be equivalently reformulated by
\begin{align}
&\mathrm{\ddot{R}}_{g,k}(\{\mathbf{f}_g\}, \gamma_{g,k}, \omega_{g,k})\\
&=
\underbrace{-{\sum}_{i=1}^{G}\mathbf{f}_i^H \mathbf{B}_{1,i,g,k}\mathbf{f}_i
+2\text{Re}\{ \mathbf{b}_{1,g,k}^H\mathbf{f}_g \}
+
c_{1,g,k}}
\limits_{\mathrm{\bar{R}}_{g,k}(\{\mathbf{f}_g\})}
.
\nonumber
\end{align}

Based on the transformation described above,
the beamformer optimization problem can be formulated as
\begin{subequations}
\begin{align}
\textrm{(P2)}:&\mathop{\textrm{max}}
\limits_{\{\mathbf{f}_g\}
}\
\bigg\{ \mathrm{R}_{s} 
= 
{\sum}_{g=1}^{G} 
\mathop{\textrm{min}}
\limits_{ k \in \mathcal{K} } 
\mathrm{\bar{R}}_{g,k}(\{\mathbf{f}_g\})
 \bigg\}\label{P2_obj}\\
\textrm{s.t.}\ 
& \mathbf{f}_{g}^H\mathbf{\bar{A}}_n\mathbf{f}_{g} \leq P_t, \forall n \in \mathcal{N}, \forall g \in \mathcal{G},\label{P2_c_1}
\end{align}
\end{subequations}

{
In the next,
to efficiently obtain the solution of the beamformer,
we iteratively update each TRIS transceiver's beamformer vector in sequence.
With other beamformer variables (i.e., $\{\mathbf{f}_{i}, i \neq g\}$) being fixed,
the function
$\mathrm{\bar{R}}_{g,k}(\{\mathbf{f}_g\})$
with respect to (w.r.t.)
the $g$-th beamformer variable $\mathbf{f}_{g}$
can be written as
\begin{align}
&\mathrm{\bar{R}}_{g,j,k}(\{\mathbf{f}_g\})\\
&= 
\underbrace{-\mathbf{f}_g^H \mathbf{B}_{1,g,j,k}\mathbf{f}_g
+2\text{Re}\{ \mathbf{\tilde{b}}_{1,g,k}^H\mathbf{f}_g \}
+
\tilde{c}_{1,g,k}}
\limits_{\mathrm{\tilde{R}}_{g,j,k}(\mathbf{f}_g)},\nonumber
\end{align}
where the newly introduced coefficients are given as
\begin{align}
&\mathbf{\tilde{b}}_{1,g,k}
\triangleq
\begin{cases}
\mathbf{{b}}_{1,g,k} , j =g \\
\mathbf{0}\in \mathbb{R}^{N \cdot K \times 1}, j\neq g
\end{cases}\\
&\tilde{c}_{1,j,k}
\triangleq
\begin{cases}
c_{1,j,k} - \sum_{i\neq g}^{G} \mathbf{f}_i^H \mathbf{B}_{1,i,g,k} \mathbf{f}_i , j =g \\
c_{1,j,k}
+ 2\text{Re}\{ \mathbf{b}_{1,j,k}^H \mathbf{f}_j \}
 - \sum_{i\neq g}^{G} \mathbf{f}_i^H \mathbf{B}_{1,i,j,k} \mathbf{f}_i, j\neq g
\end{cases}.\nonumber
\end{align}
}

{
Therefore,
the optimization problem w.r.t. the variable $\mathbf{f}_{g}$
is given as
\begin{subequations}
\begin{align}
\textrm{(P3)}:&\mathop{\textrm{max}}
\limits_{\mathbf{f}_g
}\
\bigg\{ 
\mathrm{R}_{s} 
= 
{\sum}_{j=1}^{G} 
\mathop{\textrm{min}}
\limits_{ k \in \mathcal{K} } 
\mathrm{\tilde{R}}_{g,j,k}(\mathbf{f}_g)
 \bigg\}\label{P3_obj}\\
\textrm{s.t.}\ 
& \mathbf{f}_{g}^H\mathbf{\bar{A}}_n\mathbf{f}_{g} \leq P_t, \forall n \in \mathcal{N}.\label{P3_c_1}
\end{align}
\end{subequations}
}

Besides,
we can observe that the coefficient $ \mathbf{B}_{1,j,g,k} $ is a block diagonal matrix.
In light of this and the structure of the power constraint in (\ref{P3_c_1}) for the TRIS transceiver unit,
the variable $\mathbf{f}_g$
can be decomposed into 
$N$ subvariables,
where each subvariable is defined as:
\begin{align}
\mathbf{\bar{f}}_{g,n} \triangleq & [ \mathbf{f}_g( n), \mathbf{f}_g(N + n), \cdots, \mathbf{f}_g((k-1)\times N+ n) ,\\ 
&\cdots, \mathbf{f}_g((K-1)\times N+ n) ]^T\in \mathbb{C}^{K\times 1}. \nonumber
\end{align}

Subsequently,
we denote the new notations provided in (\ref{notations_NN}).
\begin{figure*}
\begin{align}
& \mathbf{b}_{2,g,j,k,n} 
\triangleq
[\mathbf{B}_{1,g,j,k}(n,n), \mathbf{B}_{1,g,j,k}(n+N,n+N), \cdots , \mathbf{B}_{1,g,j,k}(n+(K-1) \times N,n+(K-1) \times N)  ]^T \in \mathbb{C}^{K\times 1},\label{notations_NN}\\
&\mathbf{B}_{2,g,j,k,n}
\triangleq
\text{diag}(\mathbf{b}_{2,g,j,k,n} ),
b_{3,g,j,k,n,i}
\triangleq
{\sum}_{z \neq n}^{N}
\mathbf{f}_g( (i-1)\times N + n) \mathbf{B}_{1,g,j,k}^H(n+(i-1) \times N,n+(i-1) \times N),
\nonumber\\
&
c_{3,g,j,k,n}
\triangleq
{\sum}_{i=1}^{K} {\sum}_{z \neq n}^{N}{\sum}_{v \neq n}^{N}
\mathbf{f}_g^H( (z-1)\times N + n)  \mathbf{B}_{1,g,j,k}^H(n+(z-1) \times N,n+(v-1) \times N)  \mathbf{f}_g( (v-1)\times N + n),
\nonumber\\
&\mathbf{b}_{4,g,k,n}
\triangleq
[\mathbf{\tilde{b}}_{1,g,k}(n), \mathbf{\tilde{b}}_{1,g,k}(n+N), \cdots ,\mathbf{\tilde{b}}_{1,g,k}(N+(K-1) \times N)  ]^T \in
\mathbb{C}^{K\times 1},
\nonumber\\
&\mathbf{b}_{3,g,j,k,n}
\triangleq
[ b_{3,g,j,k,n,1}, b_{3,g,j,k,n,2}, \cdots, b_{3,g,j,k,n,K} ]^T \in \mathbb{C}^{K \times 1},
c_{4,g,j,k,n}
\triangleq
{\sum}_{i \neq N}^{N} 2\text{Re}\{ \mathbf{b}_{4,j,k,n}^H \mathbf{\bar{f}}_{g,i} \},
\nonumber\\
&
\mathbf{b}_{5,g,j,k,n} 
\triangleq
\mathbf{b}_{4,g,j,k,n}
-
\mathbf{b}_{3,g,j,k,n},
c_{5,g,j,k,n}
\triangleq
\tilde{c}_{1,j,k}
-
c_{3,g,j,k,n}
+
c_{3,g,j,k,n}.\nonumber
\end{align}
\boldsymbol{\hrule}
\end{figure*}
With other subvariables (i.e., $\{  \mathbf{\bar{f}}_{g,i}, i \neq n \}$)  being fixed,
the function  w.r.t.
$\mathrm{\tilde{R}}_{j,k}(\mathbf{f}_g)$ 
is reformulated as
\begin{align}
&\mathrm{\tilde{R}}_{g,j,k}(\mathbf{f}_g)\\
&= 
\underbrace{
-\mathbf{\bar{f}}_{g,n}^H \mathbf{B}_{2,g,j,k,n}\mathbf{\bar{f}}_{g,n}
+2\text{Re}\{ \mathbf{b}_{5,g,j,k,n}^H\mathbf{\bar{f}}_{g,n} \}
+
c_{5,g,j,k,n}
}
\limits_{\mathrm{\acute{R}}_{g,j,k,n}(\mathbf{\bar{f}}_{g,n})},\nonumber
\end{align}

Therefore,
the update of the subvariable $\mathbf{\bar{f}}_{g,n}$  is  meant to solve the following problem
\begin{subequations}
\begin{align}
\textrm{(P4)}:&\mathop{\textrm{max}}
\limits_{
\mathbf{\bar{f}}_{g,n}
}\
\bigg\{ 
\mathrm{R}_{s} 
= 
{\sum}_{j=1}^{G} 
\mathop{\textrm{min}}
\limits_{ k \in \mathcal{K} } 
\mathrm{\acute{R}}_{g,j,k,n}(\mathbf{\bar{f}}_{g,n})
 \bigg\}\label{P4_obj}\\
\textrm{s.t.}\ 
& \mathbf{\bar{f}}_{g,n}^H\mathbf{\bar{f}}_{g,n} \leq P_t.\label{P4_c_1}
\end{align}
\end{subequations}

Note that the objective function of the problem (P4) is non-differentiable.
In the next,
we will use the smooth approximation theory \cite{ref_Log_sum}$-$\cite{ref_Log_sum_1} to approximate 
$ \mathop{\textrm{min}}_{ k \in \mathcal{K} } 
\{\mathrm{\acute{R}}_{g,j,k,n} (\mathbf{\bar{f}}_{g,n})\}$,
which can be given as
\begin{align}
&\mathop{\textrm{min}}_{ k \in \mathcal{K} } 
\{\mathrm{\acute{R}}_{g,j,k,n} (\mathbf{\bar{f}}_{g,n})\}
\approx
\mathrm{\breve{R}}_{g,j,n} (\mathbf{\bar{f}}_{g,n})\\
&=
-\frac{1}{\mu_{g,j,n}} \text{log}\bigg( \sum_{k \in \mathcal{K}} 
\text{exp} \big( - \mu_{g,j,n} \mathrm{\acute{R}}_{g,j,k,n} (\mathbf{\bar{f}}_{g,n})  
\big) \bigg),\nonumber
\end{align}
where
the function 
$\mathrm{\breve{R}}_{g,j,n} (\mathbf{\bar{f}}_{g,n})$
denotes 
a smooth function
and 
is the lower bound of 
$ \mathop{\textrm{min}}_{ k \in \mathcal{K} } 
\{\mathrm{\acute{R}}_{g,j,k,n} (\mathbf{\bar{f}}_{g,n})\}$,
and
$\mu_{g,j,n}$ represents the smoothing parameter
that satisfies the following inequalities:
\begin{align}
&\mathrm{\breve{R}}_{g,j,n} (\mathbf{\bar{f}}_{g,n})
 + \frac{1}{\mu_{g,j,n}} \text{log}(\vert \mathcal{K} \vert)\\
&\geq
\mathop{\textrm{min}}_{ k \in \mathcal{K} } 
\{\mathrm{\acute{R}}_{g,j,k,n} (\mathbf{\bar{f}}_{g,n})\}
\geq
\mathrm{\breve{R}}_{g,j,n} (\mathbf{\bar{f}}_{g,n}).\nonumber
\end{align}

According to the proof of \cite{ref_RIS_A_2},
the function 
$ -\frac{1}{\mu_{g,j,n}} \text{log}\big( \sum_{k \in \mathcal{K}} 
\text{exp} \big( - \mu_{g,j,n} \mathrm{\acute{R}}_{g,j,k,n} (\mathbf{\bar{f}}_{g,n})  
\big) \big)$  
is monotonically increasing and concave
w.r.t. $\mathrm{\acute{R}}_{g,j,k,n} (\mathbf{\bar{f}}_{g,n})$.
Note that 
the function 
$\mathrm{\acute{R}}_{g,j,k,n} (\mathbf{\bar{f}}_{g,n})$  
is concave in the variable $\mathbf{\bar{f}}_{g,n}$.
By leveraging the composition principle \cite{ref_Convex Optimization}, 
we can observe that the function $\mathrm{\breve{R}}_{g,j,n} (\mathbf{\bar{f}}_{g,n})$ is concave in the variable $\mathbf{\bar{f}}_{g,n}$ as well.

Once the appropriate value of $\mu_{g,j,n}$ has been determined, 
we can proceed to solve the following problem
\begin{subequations}
\begin{align}
\textrm{(P5)}:&\mathop{\textrm{max}}
\limits_{\mathbf{\bar{f}}_{g,n}
}\ 
{\sum}_{j=1}^{G} 
\mathrm{\breve{R}}_{g,j,n} (\mathbf{\bar{f}}_{g,n})
\label{P5_obj}\\
\textrm{s.t.}\
& \mathbf{\bar{f}}_{g,n}^H\mathbf{\bar{f}}_{g,n} \leq P_t.\label{P5_c_1}
\end{align}
\end{subequations}

Evidently, 
the aforementioned problem (P5) is still highly complex and poses a significant challenge to resolve.
Fortunately, this difficulty can be tackled via the MM methodology.

First,
we present a brief introduction to the MM framework \cite{ref_MM}.
The MM method addresses complex optimization problems by iteratively constructing surrogate functions for the objective and/or constraints, 
which are easier to optimize than the original ones.
Let $f(\mathbf{x})$  denote the original objective function, 
and let $\mathbb{S}_{\mathbf{x}}$ represent the feasible set, 
which is assumed to be convex w.r.t. $\mathbf{x}$. 
Denote by $\mathbf{x}^{t-1}$  the solution obtained at the $(t-1)$-th iteration. 
A surrogate function $u(\mathbf{x}  \vert\mathbf{x}^{t-1})$ of the variable $\mathbf{x}$ is then constructed 
based on the solution, i.e., $\mathbf{x}^{t-1}$, from the previous iteration.
This surrogate is optimized in place of the original objective function at each iteration.
Moreover,
the convex approximation $u(\mathbf{x} \vert\mathbf{x}^{t-1})$ is required to satisfy the following conditions
\begin{align}
&C1): u(\mathbf{x}^{t}  \vert\mathbf{x}^{t}) = f(\mathbf{x}^{t}), \forall \mathbf{x}^{t} \in  \mathbb{S}_{\mathbf{x}}; \\
&C2): f(\mathbf{x}) \geq u(\mathbf{x}  \vert\mathbf{x}^{t}), \forall \mathbf{x}^{t}, \mathbf{x} \in  \mathbb{S}_{\mathbf{x}};\nonumber\\
&C3): \nabla_{\mathbf{x}^{t}}u(\mathbf{x}^{t}  \vert\mathbf{x}^{t})  =  \nabla_{\mathbf{x}^{t}} f(\mathbf{x}^{t});\nonumber\\
&C4): u(\mathbf{x}  \vert\mathbf{x}^{t})\ \text{is continuous in}\ \mathbf{x}\ \text{and}\ \mathbf{x}^{t}.\nonumber
\end{align}

The first condition dictates that the convex approximation function $u(\mathbf{x}^{t}  \vert\mathbf{x}^{t})$
and the original function  $f(\mathbf{x}^{t})$
must have identical value at the point $\mathbf{x}^{t}$. 
The second condition is that the original function  $f(\mathbf{x})$ 
establishes a global upper bound for the convex surrogate $u(\mathbf{x}  \vert\mathbf{x}^{t})$.
Finally, 
the third condition is for the first-order derivatives of 
both the approximation and the original function to coincide.

Therefore,
following the MM framework,
a lower bound of the objective function (\ref{P5_obj}) at the point $\mathbf{\bar{f}}_{g,n,0}$
can be constructed as follows
\begin{align}
&\mathrm{\breve{R}}_{g,j,n}(\mathbf{\bar{f}}_{g,n,0})
\geq 
\mathrm{\grave{R}}_{g,j,n}(\mathbf{\bar{f}}_{g,n}\vert \mathbf{\bar{f}}_{g,n,0})\label{MM_lower_bound}\\
&= c_{6,g,j,n} + 2\text{Re}\{\mathbf{b}_{6,g,j,n}^H \mathbf{\bar{f}}_{g,n}\} + \alpha_{g,j,n}\mathbf{\bar{f}}_{g,n}^H\mathbf{\bar{f}}_{g,n}, \nonumber 
\end{align}
where
$\mathbf{\bar{f}}_{g,n,0}$ is the latest value of $\mathbf{\bar{f}}_{g,n}$,
and the newly added coefficients are given in (\ref{MM_coefficient}).
\begin{figure*}
\begin{align}
& h_{g,j,k,n}( \mathbf{\bar{f}}_{g,n,0} )
 \triangleq
\frac{\text{exp}\big( - \mu_{g,j,n} \mathrm{\acute{R}}_{g,j,k,n} (\mathbf{\bar{f}}_{g,n,0})  \big)}
{\sum_{k\in \mathcal{K}}  \text{exp} \big( - \mu_{g,j,n} \mathrm{\acute{R}}_{g,j,k,n} (\mathbf{\bar{f}}_{g,n,0})  \big)  }, 
% \label{MM_coefficient} \\
%&
\alpha_{g,j,n}
\triangleq 
- \mathop{\textrm{max}}
\limits_{ k \in \mathcal{K} } \{\lambda_{\text{max}}(\mathbf{B}_{2,g,j,k,n}) \}
-2\mu_{g,j,n}\mathop{\textrm{max}}
\limits_{ k \in \mathcal{K} }
\{ tc_{g,j,k,n}   \}
,\nonumber\\
&
tc_{g,j,k,n} 
\triangleq 
\lambda_{\text{max}}(\mathbf{B}_{2,g,j,k,n}\mathbf{B}_{2,g,j,k,n}^H)P_{t}
+ \Vert \mathbf{b}_{5,g,j,k,n} \Vert_2^2
+ 2\sqrt{P_{t}} \Vert \mathbf{B}_{2,g,j,k,n}\mathbf{b}_{5,g,j,k,n}  \Vert_2
,\nonumber\\
& \mathbf{b}_{6,g,j,n}
\triangleq 
{\sum}_{k\in \mathcal{K}}  h_{g,j,k,n}( \mathbf{\bar{f}}_{g,n,0} )
 ( \mathbf{b}_{5,g,j,k,n} - \mathbf{B}_{2,g,j,k,n}^H\mathbf{\bar{f}}_{g,n,0}  ) 
 - \alpha_{g,j,n}\mathbf{\bar{f}}_{g,n,0},\nonumber\\ 
 & c_{6,g,j,n} \triangleq \mathrm{\breve{R}}_{g,j,n}(\mathbf{\bar{f}}_{g,n,0}) 
- 2\text{Re}\{ \mathbf{b}_{7,g,j,n}^H  \mathbf{\bar{f}}_{g,n,0} \}  + \alpha_{g,j,n}\mathbf{\bar{f}}_{g,n,0}^H\mathbf{\bar{f}}_{g,n,0}.\label{MM_coefficient}
\end{align}
\boldsymbol{\hrule}
\end{figure*}
The details of their derivation can be found in Appendix A.

Based on the above MM transformation,
the objective function of problem (P5) can be replaced by (\ref{MM_lower_bound}).
Consequently,
the update of 
$ \mathbf{\bar{f}}_{g,n} $
can be achieved by optimizing a convex lower bound of the objective function (\ref{MM_lower_bound}),
which is formulated as
\begin{subequations}
\begin{align}
\textrm{(P6)}:&\mathop{\textrm{max}}
\limits_{\mathbf{\bar{f}}_{g,n}
}\ 
\bar{\alpha}_{g,n}\mathbf{\bar{f}}_{g,n}^H\mathbf{\bar{f}}_{g,n} 
+ 2\text{Re}\{\mathbf{b}_{8,g,n}^H \mathbf{\bar{f}}_{g,n}\}  + c_{7,g,n} 
\label{P6_obj}\\
&\textrm{s.t.}\
 \mathbf{\bar{f}}_{g,n}^H\mathbf{\bar{f}}_{g,n} \leq P_t.\label{P6_c_1}
\end{align}
\end{subequations}
where 
\begin{align}
&\bar{\alpha}_{g,n} \triangleq {\sum}_{j=1}^{G} \alpha_{g,j,n},
 \mathbf{b}_{8,g,n} \triangleq {\sum}_{j=1}^{G} \mathbf{b}_{6,g,j,n},\\
& c_{7,g,n} \triangleq {\sum}_{j=1}^{G} c_{6,g,j,n}.\nonumber
\end{align}

{
Note that the coefficient $\bar{\alpha}_{g,n}$ is negative.}
Thus,
the problem (P6) is convex and can be tracked by off-the-shelf numerical solvers, e.g., CVX.

However, 
the aforementioned method for solving (P6) relies on the interior point (IP) method \cite{ref_Convex Optimization}, 
which typically entails high computational complexity. 
In the following, 
by leveraging the Lagrangian multiplier method,
we propose a CVX-free solution for efficiently addressing (P6).

First,
the Lagrange function associated with the problem (P6) is formulated as
\begin{align}
\mathcal{L}(\mathbf{\bar{f}}_{g,n},\nu) =& 
 - \bar{\alpha}_{g,n}\mathbf{\bar{f}}_{g,n}^H\mathbf{\bar{f}}_{g,n} - 2\text{Re}\{\mathbf{b}_{8,g,n}^H \mathbf{\bar{f}}_{g,n}\}  \\
 &- c_{7,g,n} + \nu( \mathbf{\bar{f}}_{g,n}^H\mathbf{\bar{f}}_{g,n} - P_t ),\nonumber
\end{align}
where
$\nu$
is the Lagrangian multiplier associated with the power constraint (\ref{P6_c_1}).

Subsequently, 
we take the first-order derivative of the Lagrange function $\mathcal{L}(\mathbf{\bar{f}}_{n},\nu)$ w.r.t. 
the variable $\mathbf{\bar{f}}_{g,n}$ and equate it to zero, 
which yields:
\begin{align}
\frac{\partial \mathcal{L}(\mathbf{\bar{f}}_{g,n},\nu)}{\partial \mathbf{\bar{f}}_{g,n}} = \mathbf{0}.
\end{align}

Then,
the solution of $\mathbf{\bar{f}}_{g,n}$ can be given as
\begin{align}
\mathbf{\bar{f}}_{g,n} = \frac{\mathbf{b}_{8,g,n}}{ \nu - \bar{\alpha}_{g,n}  }. \label{P6_closed_solution}
\end{align}

Furthermore,
by incorporating the equation (\ref{P6_closed_solution}) into the power constraint (\ref{P6_c_1}),
we can obtain the following relationship:
\begin{align}
 \frac{\mathbf{b}_{8,g,n}^H\mathbf{b}_{8,g,n}}{ (\nu - \bar{\alpha}_{g,n})^2  }\leq P_t. \label{MM_closed_power}
\end{align}

We can observe that the expression on the left-hand side of (\ref{MM_closed_power}) 
is a monotonically decreasing function w.r.t. the Lagrangian multiplier $\nu$.
Next, 
by determining whether the equality sign of the inequality is achieved,
the optimal solution to problem (P11) can be classified into the following two cases:
\begin{itemize}
\item[] \underline{CASE-I}:
If the equation (\ref{MM_closed_power}) holds when $\nu = 0$, 
then the optimal solution for (P6) can be formulated as:
\begin{align}
\mathbf{\bar{f}}_{g,n}^{\star} = -\frac{\mathbf{b}_{8,g,n}}{  \bar{\alpha}_{g,n}  }. 
\end{align}

\item[] \underline{CASE-II}: 
When $\nu > 0 $,
the optimal solution of problem (P6) is given by
\begin{align}
\mathbf{\bar{f}}_{g,n}^{\star} = \sqrt{P_t}\frac{\mathbf{b}_{8,g,n}}{  \Vert\mathbf{b}_{8,g,n}\Vert_2  }. 
\end{align}

\end{itemize}

The low-complexity algorithm can be summarized in Algorithm \ref{alg:1},
where
$ \mathcal{R} (\cdot) $
denotes the objective function (\ref{P0_obj})
and
$ \mathcal{F} (\cdot) $
represents the nonlinear fixed-point iteration map of
the low-complexity algorithm in (\ref{P6_closed_solution}).

\begin{algorithm}[t]
\caption{The Low-Complexity Algorithm}
\label{alg:1}
\begin{algorithmic}[1]
\STATE {initialize}
$\{\mathbf{f}_{g}^{(0)}\}$
and
$t=0$
;
\REPEAT
\STATE update $\{\gamma_{g,k}^{(t+1)}\}$ and $\{\omega_{g,k}^{(t+1)}\}$ by (\ref{FP_auxiliary_solution_gamma}) 
and (\ref{FP_auxiliary_solution_omega}), respectively;
\FOR{$g = 1$ to $G$}
\FOR{$n = 1$ to $N$}
\STATE $\mathbf{\bar{f}}_{g,n,1} = \mathcal{F} (\mathbf{\bar{f}}_{g,n}^{(t)} )$;
\STATE $\mathbf{\bar{f}}_{g,n,2} = \mathcal{F} (\mathbf{\bar{f}}_{g,n,1} )$;
\STATE $\mathbf{j}_{1} = \mathbf{\bar{f}}_{g,n,1} - \mathbf{\bar{f}}_{g,n}^{(t)}$;
\STATE $\mathbf{j}_{2} = \mathbf{\bar{f}}_{g,n,2} - \mathbf{\bar{f}}_{g,n,1} - \mathbf{j}_{1}$;
\STATE $\tau = - \frac{\Vert \mathbf{j}_{1}\Vert_2}{\Vert\mathbf{j}_{2}\Vert_2} $;
\STATE $\mathbf{\bar{f}}_{g,n}^{(t+1)} = \mathbf{\bar{f}}_{g,n}^{(t)} - 2\tau\mathbf{j}_{1} + \tau^2\mathbf{j}_{2} $;
\STATE if $\Vert\mathbf{\bar{f}}_{g,n}^{(t+1)}\Vert_2^2 > P_t$,
  $ \mathbf{\bar{f}}_{g,n}^{(t+1)} = \sqrt{P_t}  \frac{\mathbf{\bar{f}}_{g,n}^{(t+1)}}{\Vert\mathbf{\bar{f}}_{g,n}^{(t+1)}\Vert_2}  ; 
  $
\WHILE{ $\mathcal{R} (\mathbf{\bar{f}}_{g,n}^{(t+1)}) < \mathcal{R} (\mathbf{\bar{f}}_{g,n}^{(t)})$ }
\STATE $\rho = (\rho -1)/2$;
\STATE if $\Vert\mathbf{\bar{f}}_{g,n}^{(t+1)}\Vert_2^2 > P_t$,
  $ \mathbf{\bar{f}}_{g,n}^{(t+1)} = \sqrt{P_t} \frac{\mathbf{\bar{f}}_{g,n}^{(t+1)}}{\Vert\mathbf{\bar{f}}_{g,n}^{(t+1)}\Vert_2}   $;
\ENDWHILE
\ENDFOR
\ENDFOR
%\UNTIL{$convergence$;}
\STATE $t++$;
\UNTIL{$convergence$;}
\end{algorithmic}
\end{algorithm}

\subsection{Complexity}
{
In this subsection, 
we briefly discuss the complexity of solving the problem (P6).
According to the complexity analysis in \cite{ref_complexity}, 
the computational complexity of solving the problem (P6) via SOCP solver is
$
\mathcal{O}(K^3).
$
In comparison, the closed-form solution by the Lagrangian multiplier method has negligible complexity. 
}

\section{Numerical Results}

\begin{figure}[t]
	\centering
	\includegraphics[width=.45\textwidth]{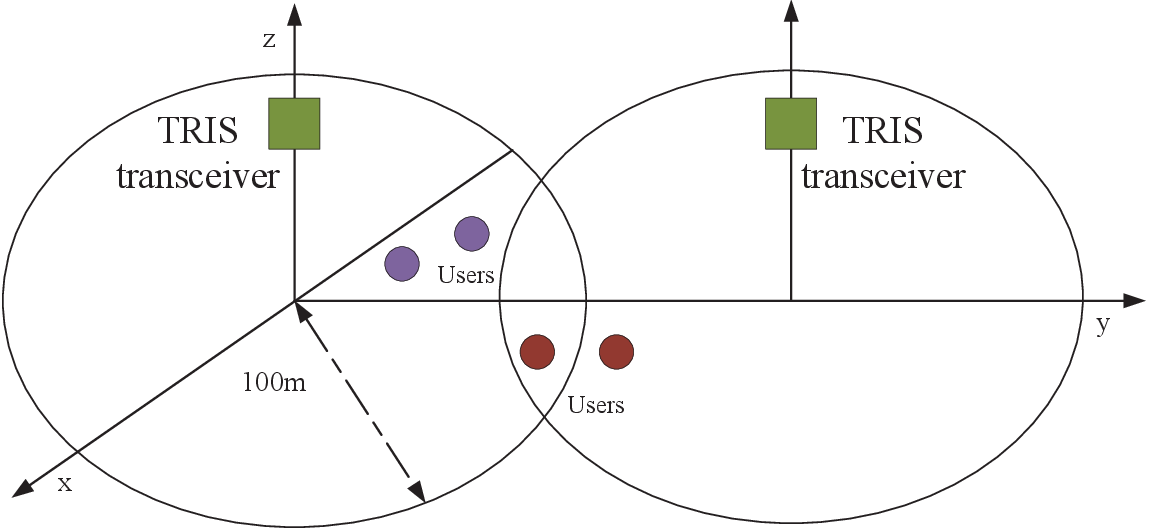}
	\caption{Simulation setup for a multi-cell MISO communication system using the TRIS transceiver.}
	\label{fig.2}
\end{figure}

In this section,
comprehensive simulation results
are presented to demonstrate the effectiveness of the low-complexity algorithm for 
the considered TRIS transceiver-enabled downlink multi-cell MISO communication system.
The simulated multi-cell communication system setting is shown in  Fig. \ref{fig.2},
which includes $G = 2$ cells, each containing one TRIS transceiver and $K = 2$ mobile users.
In the experiment, 
the first TRIS transceiver is located at the three-dimensional (3D) coordinates (0,0,4.5),
while the second TRIS transceiver is located at (140,0,4.5).
In each cell, 
all users are randomly distributed within a circle of 100m in radius, 
centered around the TRIS transceiver, and are positioned at a height of 1.5m.
The antenna spacing is set to half the wavelength of the carrier.
We assume that 
the TRIS transceiver-user link
follows the Rician distribution with a Rician factor of 5dB.
The path loss exponent of the TRIS transceiver-user link is $\alpha_{l} = 3.2$.
The maximum transmit power for each element of the TRIS transceiver is set as 10dBm. 
{The simulation parameters are summarized in Table \ref{Simulation Parameters}.}

\begin{table}[t]
%\begin{small}
\centering
\caption{{Simulation Parameters}}
{
\begin{tabular}{|c|c|} \hline \label{Simulation Parameters}
Parameter                                  & Value                  \\ \hline
G                                          & 2                      \\ \hline
K                                          & 2                      \\ \hline
Distance between two TIRS transceivers     & 140m                   \\ \hline
The maximum radius of cell                 & 100m                   \\ \hline
Rician factor                              & 5dB                    \\ \hline
Path loss exponent                         & $\alpha_{l} = 3.2$     \\ \hline
The maximum power for each unit            & 10dBm                  \\ \hline
\end{tabular}}

%\end{small}
\end{table}

\begin{figure}[t]
	\centering
	\includegraphics[width=.5\textwidth]{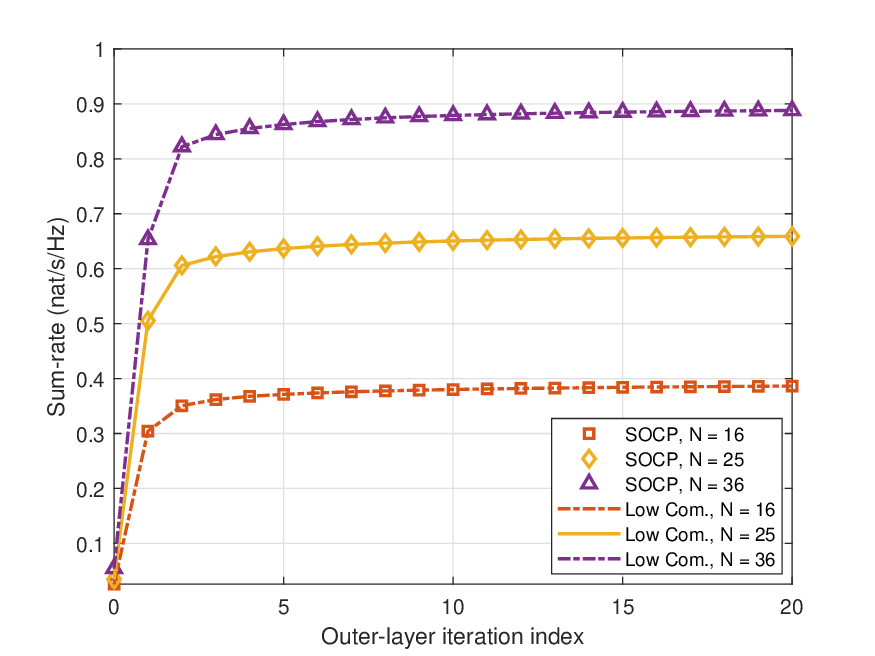}
	\caption{Convergence of Alg. 1.}
	\label{fig.3}
\end{figure}

First, we label the methods for solving problem (P6) that utilize the analytical solution and CVX as ``Low Com." and ``SOCP", respectively.
{
To ensure a fair comparison, 
both algorithm implementations are initialized at a common starting point for each channel realization.}
Fig. \ref{fig.3} shows the overall convergence behaviours of our proposed algorithms.
As depicted in the figure,
the sum-rate for both algorithms monotonically increases with the iteration index,
demonstrating the substantial gains compared to the initial point.
Both algorithms consistently deliver identical performance across all tested settings. 
Furthermore, they typically achieve convergence in under 10 iterations.
Moreover,
the sum-rate performance increases as the number of TRIS transceiver units $N$ increases.

\begin{figure}[t]
	\centering
	\includegraphics[width=.50\textwidth]{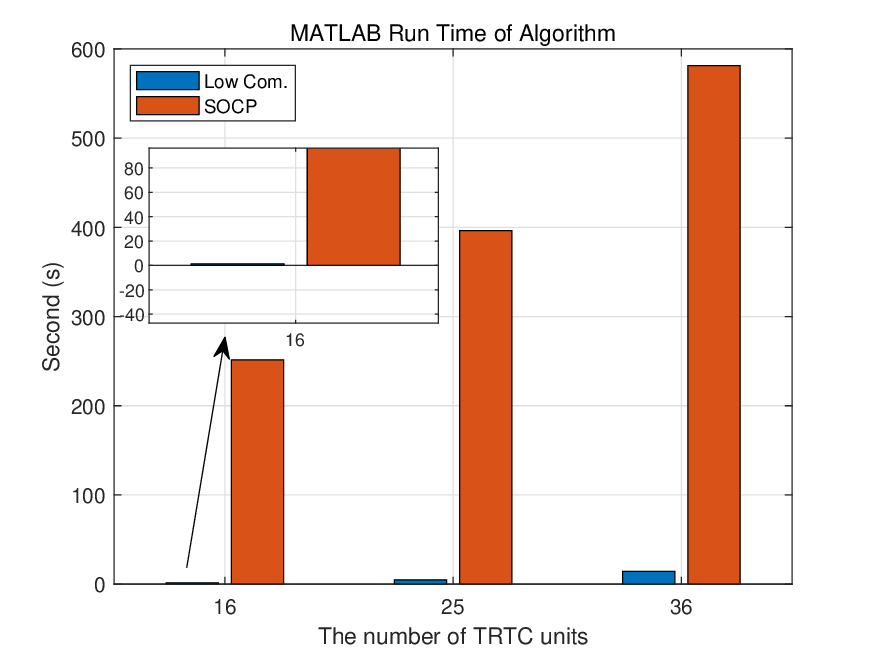}
	\caption{Comparison of MATLAB Run Time.}
	\label{fig.4}
\end{figure}

In addition,
following the convergence analysis,
we evaluate the computational complexity of the proposed algorithms.
A comparison of the MATLAB runtimes for two algorithms is provided in Fig. \ref{fig.4} across a range of TRIS transceiver element counts $N$.
The figure indicates that the ``SOCP" and ``Low Com." algorithms have the longest and shortest runtimes, respectively.
Notably, 
the ``Low Com." method is substantially more efficient,
requiring computation time that is two orders of magnitude less than that of the ``SOCP" algorithm.
{
For example,
when $N=36$, the ``Low Com." method requires only 2.5\% of the run time needed by the ``SOCP" algorithm.
}
By combining Fig. \ref{fig.3} and Fig. \ref{fig.4}, 
it is clear that the ``Low Com." method is more efficient than the ``SOCP" algorithm while still ensuring optimal sum-rate performance.

\begin{figure}[t]
	\centering
	\includegraphics[width=.50\textwidth]{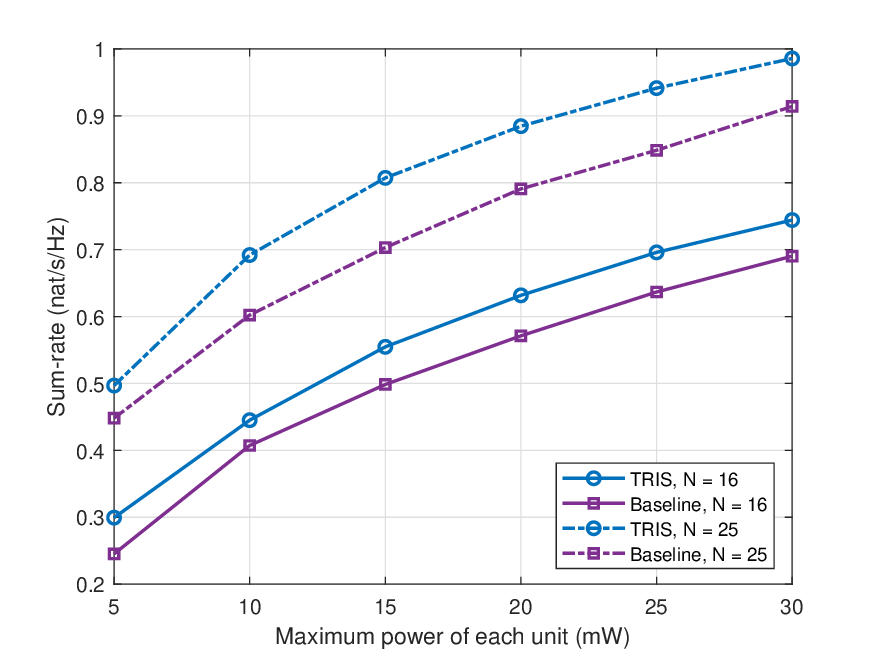}
	\caption{Sum-rate versus the maximum power of each transmissive unit.}
	\label{fig.5}
\end{figure}

{
For comparison,
we consider a baseline scheme 
that employs the traditional multi-antenna transceivers in the multi-cell system,
and the corresponding optimization problem can be written as
\begin{subequations}
\begin{align}
\textrm{(P)}:&\mathop{\textrm{max}}
\limits_{\{\mathbf{f}_g\}
}\
\bigg\{ \mathrm{R}_{s} (\{\mathbf{f}_g\}) 
= 
{\sum}_{g=1}^{G} 
\mathop{\textrm{min}}
\limits_{ k \in \mathcal{K} } 
\mathrm{R}_{g,k}(\{\mathbf{f}_g\}) 
 \bigg\}\\
\textrm{s.t.}\ 
& \mathbf{f}_{g}^H\mathbf{f}_{g} \leq N P_t, \forall g \in \mathcal{G},
\end{align}
\end{subequations}
where the above problem can still be solved by our proposed algorithm with slight modification.
}

We label the schemes that utilize the TRIS transceiver and traditional multi-antenna transceivers as ``TRIS" and ``Baseline", respectively.
Fig. \ref{fig.5} illustrates the sum-rate performance of two schemes versus 
the power budget available to each TRIS transceiver unit.
Clearly, 
the sum-rate for all schemes increases monotonically as the TRIS transceiver unit's maximum transmit power grows, 
which confirms the effectiveness of power enhancement.
Compared to the ``Baseline" scenario, 
the deployment of the TRIS transceiver significantly boosts the sum-rate.
Furthermore, 
for both ``TRIS" and ``Baseline" schemes, 
the ``N = 25" configuration achieves a significantly higher sum-rate than the ``N = 16" configuration under identical conditions.
{
For $P_t = 10$mW, 
the sum-rates of the ``TRIS'' and ``Baseline'' schemes in the $N = 25$ case reach up to 153\% and 146\% of their sum-rates in the 
$N = 16$ case, respectively.
}
Additionally,
the performance gap between ``TRIS" and ``Baseline" schemes in the ``N = 25" case is larger than observed in the ``N = 16" case.

\begin{figure}[t]
	\centering
	\includegraphics[width=.50\textwidth]{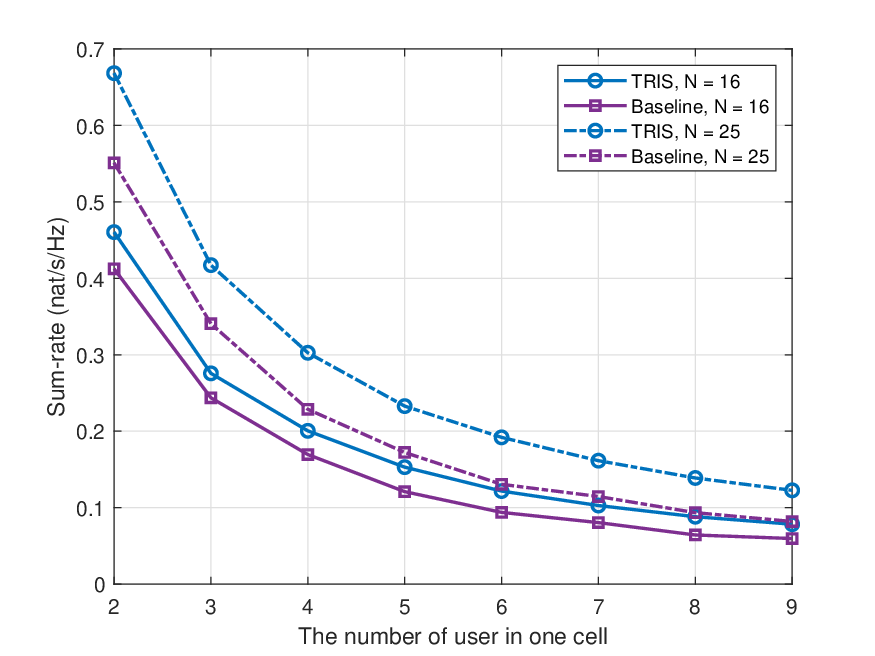}
	\caption{Sum-rate versus the number of mobile users in each cell.}
	\label{fig.6}
\end{figure}

In Fig. \ref{fig.6}, 
we illustrate the achievable sum-rate performance of the two considered schemes versus the number of users in each cell.
For all schemes, 
the sum-rate decreases monotonically as the number of users increases, 
a trend that holds under various TRIS transceiver unit settings.
Specifically, 
the total sum-rate decreases rapidly as the number of users in each cell increases from $2$ to $5$. 
In contrast, once the user count reaches $6$ and beyond, 
the rate of decline in total sum-rate slows significantly.
Moreover, 
for a fixed number of users, 
both ``TRIS" and ``Baseline" schemes achieve significantly higher sum-rates 
in the case of ``N = 25'' compared to the case of ``N = 16''.
{
For instance,
when $K=2$, the sum-rate of the ``TRIS'' scheme in the $N = 25$ case is up to 155\% of that in the $N = 16$ case, 
while the ``Baseline'' scheme reaches up to 142\%.
}
Additionally, 
as the number of users per cell increases, 
the sum-rate performance gap between the ``N = 25'' and ``N = 16'' cases narrows for both the TRIS and Baseline schemes.

\begin{figure}[t]
	\centering
	\includegraphics[width=.50\textwidth]{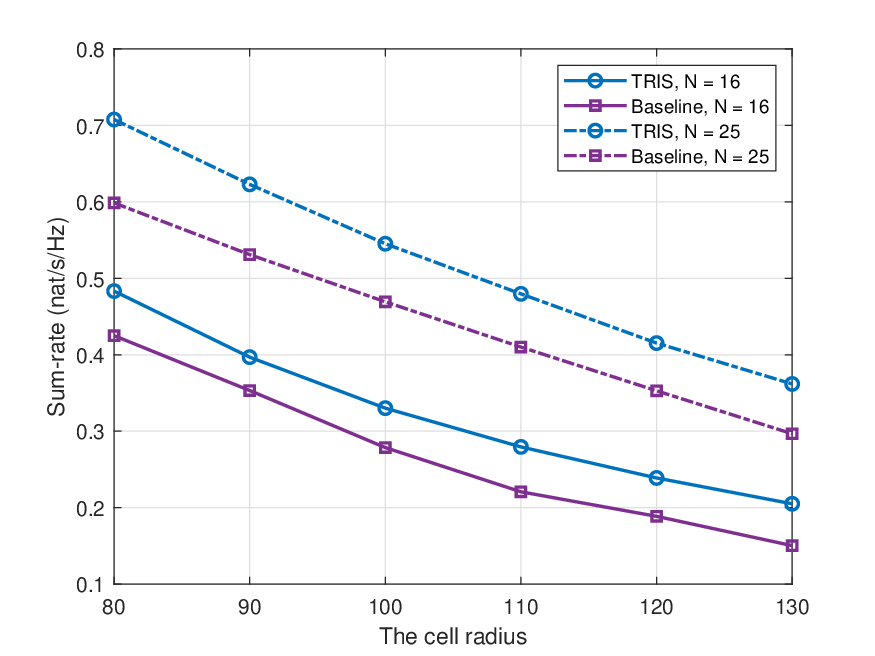}
	\caption{Sum-rate versus the cell radius.}
	\label{fig.7}
\end{figure}

Fig. \ref{fig.7} demonstrates the effect of the cell radius on the performance of all schemes.
As the cell radius is increased from $80$m to $130$m, 
a monotonic decrease in the achievable sum-rate is observed for all evaluated schemes.
Given the same system parameters, 
the two strategies exhibit a considerably enhanced sum-rate performance at N = 25 relative to N = 16.

\begin{figure}[t]
	\centering
	\includegraphics[width=.50\textwidth]{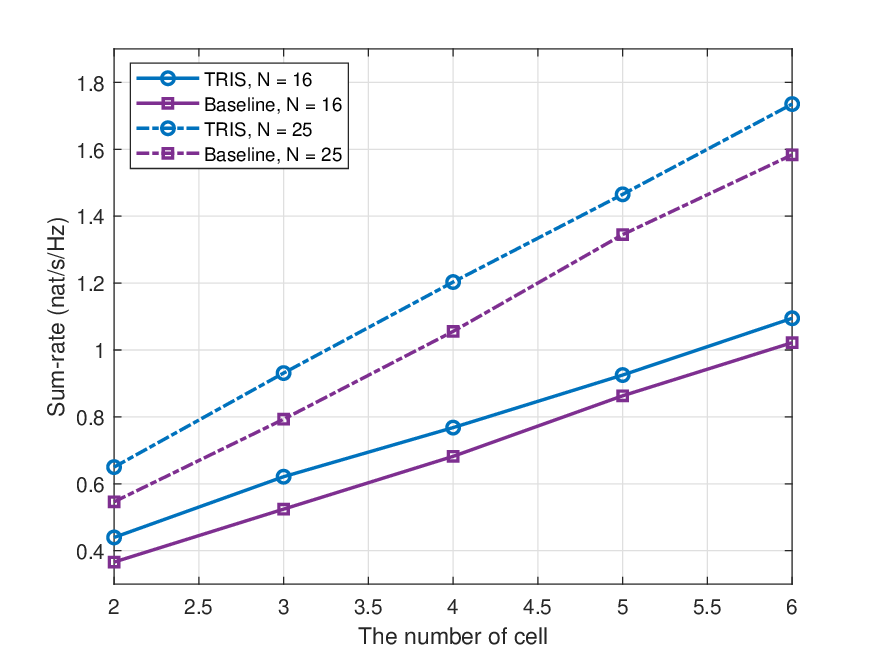}
	\caption{Sum-rate versus the number of cells.}
	\label{fig.8}
\end{figure}

The trend of sum-rate performance w.r.t. the number of cells $G$ is illustrated in Fig. \ref{fig.8}. 
It can be observed that as the number of cells increases, 
the sum-rate performance improves for both the ``TRIS" and ``Baseline" scenarios. 
Furthermore, the novel TRIS transceiver consistently outperforms conventional transceivers. 
Under the same system settings, 
the sum-rate performance of all schemes is significantly higher at 
N = 25 compared to N = 16, 
and the performance gap between these cases widens as the number of cells increases.
{
When $G = 2$, 
the sum-rate of the ``TRIS'' and ``Baseline'' schemes in the $N = 25$ scenario 
are up to 148\% and 149\% of their respective sum-rates in the $N = 16$ scenario.
}

\begin{figure}[t]
	\centering
	\includegraphics[width=.50\textwidth]{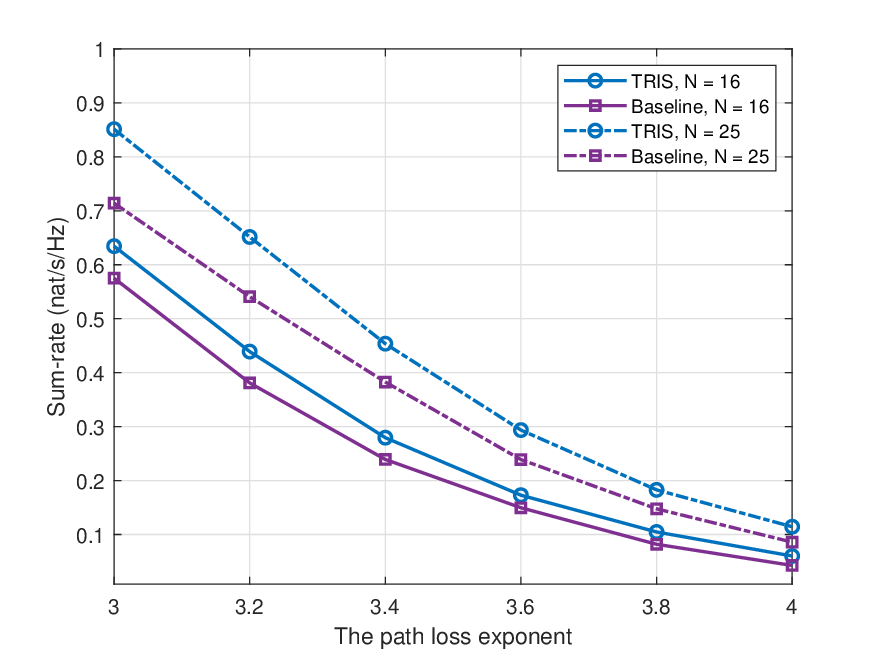}
	\caption{Sum-rate versus the path loss exponent.}
	\label{fig.9}
\end{figure}

Fig. \ref{fig.9} examines the effect of the path loss exponent of the TRIS transceiver-user channel on the sum-rate.
When the path loss exponent grows from $3.0$ to $4.0$,
the sum-rate attained by all considered schemes declines monotonically and consistently.
Furthermore, 
it is observed that the rate performance gap between the ``TRIS" and ``Baseline" scenarios progressively lessens as the path loss exponent increases.
Furthermore, 
the sum-rate performance of all schemes is significantly improved when the number of TRIS transceiver elements is increased from $16$ to $25$.
However, 
it is also noteworthy that the sum rate gap between ``N = 16'' and ``N = 25'' cases also decreases as the path loss exponent increases.

\begin{figure}[t]
	\centering
	\includegraphics[width=.50\textwidth]{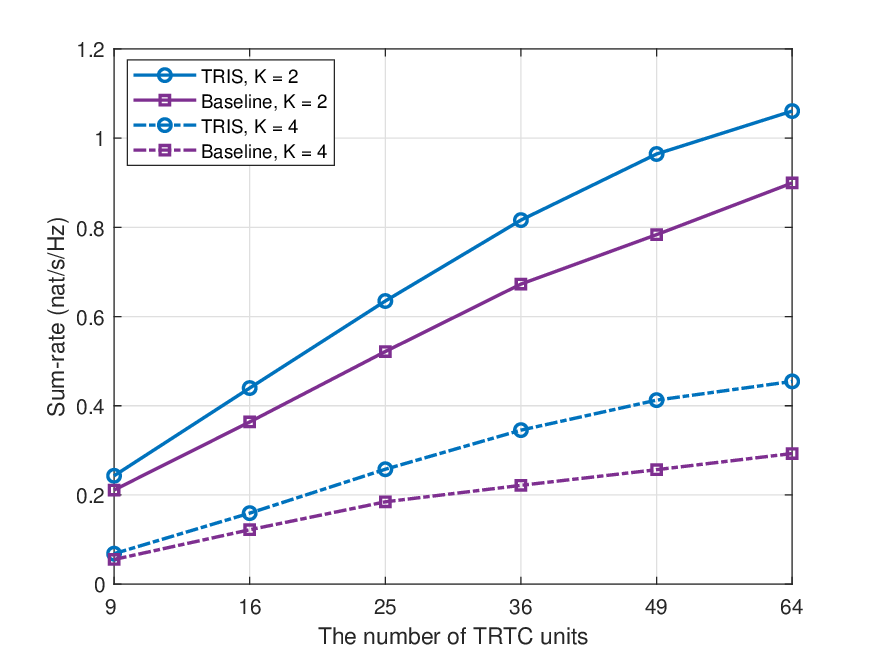}
	\caption{Sum-rate versus the number of TRIS transceiver units.}
	\label{fig.10}
\end{figure}

Fig. \ref{fig.10} depicts the effect of the number of TRIS transceiver elements. 
It is evident that increasing the number of elements enhances the beamforming gain for all schemes,
which is due to the fact that more TRIS transceiver elements provide higher diversity gain. 
Moreover, the rate of increase in sum-rate w.r.t. $N$  is significantly lower for $K=2$ compared to $K=4$.
Besides, 
it is observed that the rate performance gap between the ``TRIS" and ``Baseline" scenarios 
gradually increases as the number of TRIS transceiver elements increases.

\section{Conclusions}
%\vspace{0.2 cm}

This paper studies a TRIS transceiver-enabled multi-cell MISO communication system 
with the objective of maximizing the minimum rate across all cells to ensure rate fairness 
by optimizing the transmit beamforming vectors at the TRIS transceivers, 
while satisfying the transmit power limits of individual TRIS transceiver units.
To solve the challenging max-min rate optimization problem,
we develop an efficient and fully analytic-based solution, 
which does not depend on any numerical solvers and has low complexity,
by employing the MM methodology combined with a smooth approximation technique.
The numerical results reveal that the proposed optimization approaches greatly improve sum-rate performance 
and validate the potential of the TRIS transceiver as a novel transceiver solution for wireless networks 
emphasizing low cost and minimal power consumption.
In addition, 
the computational complexity of the proposed algorithm is much less than that of the method relying on the numerical solvers, e.g., CVX.

\appendix
\subsection{Proof of (\ref{MM_lower_bound})}
\normalem
Proof:
It is noteworthy that the function 
$\mathrm{\breve{R}}_{g,j,n} (\mathbf{\bar{f}}_{g,n})$
is twice differentiable and concave w.r.t. the variable $\mathbf{\bar{f}}_{g,n}$.
This indicates that the second derivatives of the function exist and are continuous. 

Therefore,
by combining the characteristic of the function $\mathrm{\breve{R}}_{g,j,n} (\mathbf{\bar{f}}_{g,n})$
with the MM methodology,
we can construct a quadratic surrogate function to minorize $\mathrm{\breve{R}}_{g,j,n} (\mathbf{\bar{f}}_{g,n})$,
which can be formulated as follows
\begin{align}
&\mathrm{\breve{R}}_{g,j,n} (\mathbf{\bar{f}}_{g,n})\geq
\mathrm{ \grave{R} }_{g,j,n}\label{Appendix_A_surrogate function} \\
&\triangleq
\mathrm{\breve{R}}_{g,j,n} (\mathbf{\bar{f}}_{g,n,0})
+ 2\text{Re}\{ \mathbf{b}_{7,g,j,n}^H ( \mathbf{\bar{f}}_{g,n} - \mathbf{\bar{f}}_{g,n,0} ) \}\nonumber \\
&+ ( \mathbf{\bar{f}}_{g,n} - \mathbf{\bar{f}}_{g,n,0} )^H \mathbf{D}_{g,j,n} ( \mathbf{\bar{f}}_{g,n} - \mathbf{\bar{f}}_{g,n,0} ),\nonumber
\end{align}
where
$\mathbf{b}_{7,g,j,n} \in \mathbb{C}^{K \times 1} $
and 
$\mathbf{D}_{g,j,n} \in \mathbb{C}^{K \times K}$.

Next,
by utilizing the requirement that
the surrogate function $\mathrm{ \grave{R} }_{g,j,n}$ should satisfy the conditions $C1)-C4)$ of the MM method,
we can derive the coefficients $\mathbf{b}_{7,g,j,n}$
and 
$\mathbf{D}_{g,j,n}$, 
respectively.

Obviously,
we can find that both conditions $C1)$ and $C4)$ are already met.
In the next,
we will confirm that conditions $C3)$ and $C2)$ hold in that order.

First,
following the direction $\mathbf{ \tilde{f} }_{g,n} - \mathbf{\bar{f}}_{g,n,0}  $,
we can obtain 
the  directional derivative of the function $\mathrm{ \breve{R} }_{g,j,n}$ 
at the point $\mathbf{\bar{f}}_{g,n,0}$,
which is given by
\begin{align}
&2\text{Re}\big\{
\big( {\sum}_{k \in \mathcal{K} } h_{g,j,k,n}(\mathbf{\bar{f}}_{g,n,0} )
( \mathbf{b}_{5,g,j,k,n}^H  \label{Appendix_A_C2_1}\\
& - \mathbf{\bar{f}}_{g,n,0}^H \mathbf{\bar{B}}_{2,g,j,k,n} )   
\big)
(\mathbf{ \tilde{f} }_{g,n} - \mathbf{\bar{f}}_{g,n,0})
\big\}.\nonumber
\end{align}
where the vector $\mathbf{ \tilde{f} }_{g,n}$ belongs to $\mathbb{S}_{\mathbf{f}}$.

According to the inequality (\ref{Appendix_A_surrogate function}),
the function $\mathrm{ \grave{R} }_{g,j,n}$ of the directional derivative with direction $\mathbf{ \tilde{f} }_{g,n} - \mathbf{\bar{f}}_{g,n,0}  $ can be formulated by
\begin{align}
2\text{Re}\{ \mathbf{b}_{7,g,j,n}^H (\mathbf{ \tilde{f} }_{g,n} - \mathbf{\bar{f}}_{g,n,0}) \}.\label{Appendix_A_C2_2}
\end{align}

To satisfy condition $C3)$,
the equality of the directional derivatives found in (\ref{Appendix_A_C2_1}) and (\ref{Appendix_A_C2_2}) 
is a necessary requirement. 
Therefore, 
the following equality should hold:
\begin{align}
\mathbf{b}_{7,g,j,n}
=
 {\sum}_{k \in \mathcal{K} } h_{g,j,k,n}(\mathbf{\bar{f}}_{g,n,0} )
( \mathbf{b}_{5,g,j,k,n}
 -\mathbf{\bar{B}}_{2,g,j,k,n}^H \mathbf{\bar{f}}_{g,n,0}  )   
\big.
\end{align}

In the next,
we proceed to ensure that condition $C2)$ holds.
Furthermore,
if the surrogate function 
$\mathrm{ \grave{R} }_{g,j,n}(\mathbf{\bar{f}}_{g,n} \vert\mathbf{\bar{f}}_{g,n,0} )$
provides a lower bound for all linear segments in any direction,
the condition $C2)$ is fulfilled.
As a result, 
the following expression should hold true
\begin{align}
&\mathrm{\breve{R}}_{g,j,n} 
\big(\mathbf{\bar{f}}_{g,n,0} + \tau( \mathbf{ \tilde{f} }_{g,n} - \mathbf{\bar{f}}_{g,n,0} )\big )\label{Appendix_A_C3_1} \\
&\geq
\mathrm{\breve{R}}_{g,j,n} (\mathbf{\bar{f}}_{g,n,0} ) 
+ 2\tau\text{Re}\{ \mathbf{b}_{7,g,j,n}^H ( \mathbf{\bar{f}}_{g,n} - \mathbf{\bar{f}}_{g,n,0} ) \} \nonumber \\
&+ \tau^2( \mathbf{\bar{f}}_{g,n} - \mathbf{\bar{f}}_{g,n,0} )^H \mathbf{D}_{g,j,n} 
( \mathbf{\bar{f}}_{g,n} - \mathbf{\bar{f}}_{g,n,0} ),\nonumber
\end{align}
where
$\mathbf{\bar{f}}_{g,n} = \mathbf{\bar{f}}_{g,n,0} + \tau( \mathbf{ \tilde{f} }_{g,n} - \mathbf{\bar{f}}_{g,n,0} )$, 
$\forall \tau \in [0,1]$.

Let
$P_{g,j,n}(\tau) \triangleq \mathrm{\breve{R}}_{g,j,n} (\mathbf{\bar{f}}_{g,n,0} + \tau( \mathbf{ \tilde{f} }_{g,n}
 - \mathbf{\bar{f}}_{g,n,0} ))$
and
$p_{g,j,n,k}(\tau) \triangleq  \mathrm{\acute{R}}_{g,j,k,n} (\mathbf{\bar{f}}_{g,n,0} + \tau( \mathbf{ \tilde{f} }_{g,n}
 - \mathbf{\bar{f}}_{g,n,0} ))   $.
And then,
a sufficient condition of (\ref{Appendix_A_C3_1}) can be formulated as
\begin{align}
&\frac{ \partial^2 P_{g,j,n}(\tau)  }{\partial  \tau^2} 
\geq 
2( \mathbf{\bar{f}}_{g,n} - \mathbf{\bar{f}}_{g,n,0} )^H 
\mathbf{D}_{g,j,n} 
( \mathbf{\bar{f}}_{g,n} - \mathbf{\bar{f}}_{g,n,0} ).
\label{Appendix_A_C3_1_1}
\end{align}

First,
we obtain the first-order derivative of $P_{g,j,n}(\tau)$,
which can be expressed as
\begin{align}
\frac{ \partial P_{g,j,n}(\tau)  }{\partial  \tau} 
=
\sum_{k \in \mathcal{K}} h_{1,g,j,k,n}(\tau) \nabla_{\tau} p_{g,j,k,n}(\tau),\label{Appendix_A_C3_3}
\end{align}
where the new coefficients are defined as
\begin{align}
&h_{1,g,j,k,n}(\tau) \triangleq \frac{\text{exp}( -\mu_{g,j,n}p_{g,j,k,n}(\tau) )}
{ \sum_{k\in \mathcal{K}}  \text{exp}( -\mu_{g,j,n}p_{g,j,k,n}(\tau) ) },\\
&\nabla_{\tau} p_{g,j,k,n}(\tau) 
\triangleq
-2\tau ( \mathbf{ \tilde{f} }_{g,n} - \mathbf{\bar{f}}_{g,n,0} )^H
\mathbf{\bar{B} }_{g,j,k,n}
( \mathbf{ \tilde{f} }_{g,n} - \mathbf{\bar{f}}_{g,n,0} )\nonumber\\
&+ 2\text{Re}\{ \mathbf{b}_{5,g,j,k,n}^H( \mathbf{ \tilde{f} }_{g,n} - \mathbf{\bar{f}}_{g,n,0} ) 
- \mathbf{\bar{f}}_{g,n,0}^H  \mathbf{\bar{B} }_{g,j,k,n}
( \mathbf{ \tilde{f} }_{g,n} - \mathbf{\bar{f}}_{g,n,0} ) \}
\nonumber\\
& = 2\text{Re}\{ \mathbf{d}_{g,j,k,n}^H \mathbf{ \hat{f} }_{g,n}\},
\nonumber\\
& 
\mathbf{d}_{g,j,k,n}
\triangleq
\mathbf{b}_{5,g,j,k,n}-\mathbf{\bar{B} }_{g,j,k,n}^H(\mathbf{\bar{f}}_{g,n,0} + \tau( \mathbf{ \tilde{f} }_{g,n} - \mathbf{\bar{f}}_{g,n,0} )  ),\nonumber\\
&\mathbf{ \hat{f} }_{g,n}
\triangleq
\mathbf{ \tilde{f} }_{g,n} - \mathbf{\bar{f}}_{g,n,0}.\nonumber
\end{align}

Next,
the second-order derivative of  $P_{g,j,n}(\tau)$ 
is formulated in (\ref{Appendix_A_C3_2}),
\begin{figure*}
\begin{align}
&\frac{ \partial^2 P_{g,n}(\tau)  }{\partial  \tau^2} 
=
{\sum}_{k\in \mathcal{K}}\big(  h_{1,g,j,k,n}(\tau)  \nabla_{\tau} p_{g,j,k,n}(\tau) \label{Appendix_A_C3_2}\\
&- \mu_{g,j,n}h_{1,g,j,k,n}(\tau)  (\nabla_{\tau} p_{g,j,k,n}(\tau)  )^2
 \big)+  \mu_{g,j,n}\big( {\sum}_{k\in \mathcal{K}} h_{1,g,j,k,n}(\tau)\nabla_{\tau} p_{g,j,k,n}(\tau)  \big)^2.
 \nonumber
\end{align}
\boldsymbol{\hrule}
\end{figure*}
where 
\begin{align}
\nabla_{\tau}^2 p_{g,j,k,n}(\tau) 
&=
-2( \mathbf{ \tilde{f} }_{g,n} - \mathbf{\bar{f}}_{g,n,0} )^H
\mathbf{\bar{B} }_{g,j,k,n}
( \mathbf{ \tilde{f} }_{g,n} - \mathbf{\bar{f}}_{g,n,0} )\nonumber\\
&=-2  \mathbf{ \hat{f} }_{g,n}^H \mathbf{\bar{B} }_{g,j,k,n}\mathbf{ \hat{f} }_{g,n},\label{Appendix_A_C3_4}
\end{align}

By combining the equations (\ref{Appendix_A_C3_3})$-$(\ref{Appendix_A_C3_4}),
we can rewrite
the second-order derivative $\frac{ \partial^2 P_{g,j,n}(\tau)  }{\partial  \tau^2} $
as follows
\begin{align}
\frac{ \partial^2 P_{g,j,n}(\tau)  }{\partial  \tau^2}
=
\begin{bmatrix}
\mathbf{ \hat{f} }_{g,n}^H & \mathbf{ \hat{f} }_{g,n}^T
\end{bmatrix}
\boldsymbol{\Psi}_{g,j,n}
\begin{bmatrix}
\mathbf{ \hat{f} }_{g,n} \\
\mathbf{ \hat{f} }_{g,n}^{\ast}
\end{bmatrix},
\end{align}
with
the coefficient 
$\boldsymbol{\Psi}_{g,j,n}$ given in
(\ref{Appendix_A_C3_5}).
\begin{figure*}
\begin{align}
\boldsymbol{\Psi}_{g,j,n}
&\triangleq
\sum_{k \in \mathcal{K}}
\bigg(
h_{1,g,j,k,n}(\tau) 
\begin{bmatrix}
- \mathbf{ \bar{B}}_{g,j,k,n} & \mathbf{0} \\ 
       \mathbf{0}        & - \mathbf{ \bar{B}}_{g,j,k,n}
\end{bmatrix}
-
\mu_{g,j,n}h_{1,g,j,k,n}(\tau) 
\begin{bmatrix}
 \mathbf{ d}_{g,j,k,n}\\ 
 \mathbf{ d}_{g,j,k,n}^{\ast}
\end{bmatrix}
\begin{bmatrix}
 \mathbf{ d}_{g,j,k,n}\\ 
 \mathbf{ d}_{g,j,k,n}^{\ast}
\end{bmatrix}^H
\bigg)
\label{Appendix_A_C3_5}\\
&+\mu_{g,j,n}
\begin{bmatrix}
\sum_{k \in \mathcal{K}} h_{1,g,j,k,n}(\tau)    \mathbf{ d}_{g,j,k,n}\\ 
\sum_{k \in \mathcal{K}} h_{1,g,j,k,n}(\tau)    \mathbf{ d}_{g,j,k,n}^{\ast}
\end{bmatrix}
\begin{bmatrix}
\sum_{k \in \mathcal{K}} h_{1,g,j,k,n}(\tau)    \mathbf{ d}_{g,j,k,n}\\ 
\sum_{k \in \mathcal{K}} h_{1,g,j,k,n}(\tau)    \mathbf{ d}_{g,j,k,n}^{\ast}
\end{bmatrix}^H. \nonumber 
\end{align}
\boldsymbol{\hrule}
\end{figure*}

Similarly,
we again rewrite
the right  of the inequality (\ref{Appendix_A_C3_1_1})  as follows:
\begin{align}
&2( \mathbf{\bar{f}}_{g,n} - \mathbf{\bar{f}}_{g,n,0} )^H \mathbf{D}_{g,j,n}
 ( \mathbf{\bar{f}}_{g,n} - \mathbf{\bar{f}}_{g,n,0} )\\
&=
\begin{bmatrix}
\mathbf{ \hat{f} }_{g,n}^H & \mathbf{ \hat{f} }_{g,n}^T
\end{bmatrix}
\begin{bmatrix}
 \mathbf{D}_{g,j,n} & \mathbf{0} \\
\mathbf{0}  &   \mathbf{D}_{g,j,n}
\end{bmatrix}
\begin{bmatrix}
\mathbf{ \hat{f} }_{g,n} \\
\mathbf{ \hat{f} }_{g,n}^{\ast}
\end{bmatrix}.\nonumber
\end{align}

To satisfy condition $C2)$,
we have
\begin{align}
\boldsymbol{\Psi}_{g,j,n}
\succeq
\begin{bmatrix}
 \mathbf{D}_{g,j,n} & \mathbf{0} \\
\mathbf{0}  &  \mathbf{D}_{g,j,n}
\end{bmatrix}.
\end{align}

Therefore,
we can determine
\begin{align}
\mathbf{D}_{g,j,n} = \alpha_{g,j,n} \mathbf{I} = \lambda_{\text{min}}(\boldsymbol{\Psi}_{g,j,n})\mathbf{I} .
\end{align}

And then,
the function $\mathrm{ \grave{R} }_{g,j,n}$ in (\ref{Appendix_A_surrogate function})
can be given as
\begin{align}
&\mathrm{ \grave{R} }_{g,j,n}
=
\mathrm{\breve{R}}_{g,j,n} (\mathbf{\bar{f}}_{g,n,0})
+ 2\text{Re}\{ \mathbf{b}_{7,g,j,n}^H ( \mathbf{\bar{f}}_{g,n} - \mathbf{\bar{f}}_{g,n,0} ) \} \\
&+ ( \mathbf{\bar{f}}_{g,n} - \mathbf{\bar{f}}_{g,n,0} )^H \mathbf{D}_{g,j,n} 
( \mathbf{\bar{f}}_{g,n} - \mathbf{\bar{f}}_{g,n,0} )\nonumber\\
&= c_{6,g,j,n} + 2\text{Re}\{\mathbf{b}_{6,g,j,n}^H \mathbf{\bar{f}}_{g,n}\} + \alpha_{g,j,n}\mathbf{\bar{f}}_{g,n}^H\mathbf{\bar{f}}_{g,n},\nonumber
\end{align}
where
$c_{6,g,j,n}$
and
$\mathbf{b}_{6,g,j,n}$
are given in (\ref{MM_coefficient}).

However,
we can find that the matrix $\boldsymbol{\Psi}_{g,j,n}$ is difficult to obtain its explicit value.
Therefore,
to obtain the value of $\boldsymbol{\Psi}_{g,j,n}$,
we refer to the following lemmas,
which are given as

\begin{itemize}
\item[] {a1)}:
When both matrices $\mathbf{A}$ and $\mathbf{B}$ are Hermitian,
the following inequality holds:
\begin{align}
\lambda_{\text{min}}(\mathbf{A}) + \lambda_{\text{min}}(\mathbf{B}) \leq  \lambda_{\text{min}}(\mathbf{A}+\mathbf{B});
\end{align}

\item[] {a2)}: 
When the rank of  the matrix $\mathbf{A}$ is one,
we have
\begin{align}
\lambda_{\text{max}}(\mathbf{A}) = \text{Tr}(\mathbf{A}),
\lambda_{\text{min}}(\mathbf{A}) = 0;
\end{align}

\item[] {a3)}: 
When
$a_k, b_k \geq 0$
and
$\sum_{k=1}^{K} a_k = 1$,
we obtain
\begin{align}
 {\sum}_{k=1}^{K}a_kb_k \leq \text{max}_{k=1}^{K}b_k;
\end{align}

\item[] {a4)}: 
 $\mathbf{A}$ and $\mathbf{B}$ denote positive semidefinite matrices
and 
 $\mathbf{A}$ have maximum eigenvalue $\lambda_{\text{max}}(\mathbf{A})$. 
Then the following inequality holds:
\begin{align}
 \text{Tr}(\mathbf{A}\mathbf{B}) \leq \lambda_{\text{max}}(\mathbf{A})\text{Tr}(\mathbf{B}).
\end{align}
\end{itemize}

Next,
by utilizing the lemmas $a1)  - {a4)}$,
the lower bound of $\alpha_{g,j,n}$ is formulated in (\ref{Appendix_A_C3_6}).
\begin{figure*}
%\begin{small}
\begin{align}
\lambda_{\text{min}}(\boldsymbol{\Psi}_{g,j,n})
&\overset{ \text{a1)} }
\geq
\sum_{k \in \mathcal{K}}
h_{1,g,j,k,n}(\tau) 
\lambda_{\text{max}}
\bigg(
\begin{bmatrix}
- \mathbf{ \bar{B}}_{g,j,k,n} & \mathbf{0} \label{Appendix_A_C3_6} \\ 
        \mathbf{0}        & - \mathbf{ \bar{B}}_{g,j,k,n}
\end{bmatrix}
\bigg)\\
&-
\sum_{k \in \mathcal{K}}
\mu_{g,j,n}h_{1,g,j,k,n}(\tau) 
\lambda_{\text{max}}
\bigg(
\begin{bmatrix}
 \mathbf{ d}_{g,j,k,n}\\ 
 \mathbf{ d}_{g,j,k,n}^{\ast}
\end{bmatrix}
\begin{bmatrix}
 \mathbf{ d}_{g,j,k,n}\\ 
 \mathbf{ d}_{g,j,k,n}^{\ast}
\end{bmatrix}^H
\bigg)
\nonumber\\
&+
\mu_{g,j,n}
\lambda_{\text{min}}
\bigg(
\begin{bmatrix}
\sum_{k \in \mathcal{K}} h_{1,g,j,k,n}(\tau)   \mathbf{ d}_{g,j,k,n}\\ 
\sum_{k \in \mathcal{K}} h_{1,g,j,k,n}(\tau)   \mathbf{ d}_{g,j,k,n}^{\ast}
\end{bmatrix}
\begin{bmatrix}
\sum_{k \in \mathcal{K}} h_{1,g,j,k,n}(\tau)    \mathbf{ d}_{g,j,k,n}\\ 
\sum_{k \in \mathcal{K}} h_{1,g,j,k,n}(\tau)   \mathbf{ d}_{g,j,k,n}^{\ast}
\end{bmatrix}^H
\bigg)\nonumber\\
&
\overset{ \text{a2)} }
=
-
\sum_{k \in \mathcal{K}}
h_{1,g,j,k,n}(\tau)
\big(
\lambda_{\text{max}}(\mathbf{ \bar{B}}_{g,j,k,n})
+
2\mu_{g,j,n}
\mathbf{ d}_{g,j,k,n}^H\mathbf{ d}_{g,j,k,n}
\big)\nonumber\\
&\overset{ \text{a3)} }
\geq
-
\underset{k \in \mathcal{K}}
{\text{max}}
\{ \lambda_{\text{max}}(\mathbf{ \bar{B}}_{g,j,k,n}) \}
-
2\mu_{g,j,n}
\underset{k \in \mathcal{K}}
{\text{max}}
\{ 
\Vert
\mathbf{ d}_{g,j,k,n}
\Vert_2^2
\}.
\nonumber
\end{align}
%\end{small}
\boldsymbol{\hrule}
\end{figure*}

It should be noted that the value of 
$\Vert
\mathbf{ d}_{g,j,k,n}
\Vert_2^2
$
in (\ref{Appendix_A_C3_6}) remains difficult to obtain.
Therefore,
we turn to find its upper bound instead of the original value.
Since 
$\mathbf{\bar{f}}_{g,n} = \mathbf{\bar{f}}_{g,n,0} + \tau( \mathbf{ \tilde{f} }_{g,n} - \mathbf{\bar{f}}_{g,n,0} )$,
$\forall \tau \in [0,1]$,
the following inequality can be achieved:
\begin{align}
\Vert \mathbf{\bar{f}}_{g,n}\Vert_2^2 = 
\Vert \mathbf{\bar{f}}_{g,n,0} + \tau( \mathbf{ \tilde{f} }_{g,n} - \mathbf{\bar{f}}_{g,n,0} ) \Vert_2^2
\leq P_t.
\end{align}

And then,
an upper bound of 
the term 
$\Vert
\mathbf{ d}_{g,j,k,n}
\Vert_2^2
$
can be obtained via the lemma $a4)$,
which is given in (\ref{Appendix_A_C3_7}).
\begin{figure*}
%\begin{small}
\begin{align}
&\Vert
\mathbf{ d}_{g,j,k,n}
\Vert_2^2
=
\Vert
\mathbf{b}_{5,g,j,k,n}-\mathbf{\bar{B} }_{g,j,k,n}^H
(\mathbf{\bar{f}}_{g,n,0} + \tau( \mathbf{ \tilde{f} }_{g,n} - \mathbf{\bar{f}}_{g,n,0} )  )
\Vert_2^2\label{Appendix_A_C3_7}\\
&=
\Vert
\mathbf{b}_{5,g,j,k,n}
\Vert_2^2
+
\Vert
\mathbf{\bar{B} }_{g,j,k,n}^H(\mathbf{\bar{f}}_{g,n,0} 
+ \tau( \mathbf{ \tilde{f} }_{g,n} - \mathbf{\bar{f}}_{g,n,0} )  )
\Vert_2^2
-
2\text{Re}\{
\mathbf{b}_{5,g,j,k,n}^H
\mathbf{\bar{B} }_{g,j,k,n}^H(\mathbf{\bar{f}}_{g,n,0} + \gamma( \mathbf{ \tilde{f} }_{g,n} - \mathbf{\bar{f}}_{g,n,0} )  )
\}\nonumber\\
&
\overset{a4)}
\leq
\lambda_{\text{max}}(\mathbf{\bar{B} }_{g,j,k,n}\mathbf{\bar{B} }_{g,j,k,n}^H)
\Vert
\mathbf{\bar{f}}_{g,n,0} + \gamma( \mathbf{ \tilde{f} }_{g,n} - \mathbf{\bar{f}}_{g,n,0} )  
\Vert_2^2
+\Vert
\mathbf{b}_{5,g,j,k,n}
\Vert_2^2
-
2\text{Re}\{
\mathbf{b}_{5,g,j,k,n}^H
\mathbf{\bar{B} }_{g,j,k,n}^H(\mathbf{\bar{f}}_{g,n,0} + \gamma( \mathbf{ \tilde{f} }_{g,n} - \mathbf{\bar{f}}_{g,n,0} )  )
\}
\nonumber\\
&
\leq
\lambda_{\text{max}}(\mathbf{\bar{B} }_{g,j,k,n}\mathbf{\bar{B} }_{g,j,k,n}^H)
P_t
+\Vert
\mathbf{b}_{5,g,j,k,n}
\Vert_2^2
+
2\sqrt{P_t}\Vert\mathbf{\bar{B} }_{g,j,k,n}\mathbf{b}_{5,g,j,k,n}\Vert_2.\nonumber
\end{align}
%\end{small}
\boldsymbol{\hrule}
\end{figure*}
Specifically,
to find the last term  $2\sqrt{P_t}\Vert\mathbf{\bar{B} }_{g,j,k,n}\mathbf{b}_{5,g,j,k,n}\Vert_2$
 of the final inequality in (\ref{Appendix_A_C3_7}), 
 we solve the following optimization problem to determine its optimal solution:
\begin{subequations}
\begin{align}
\mathop{\textrm{min}}
\limits_{\mathbf{x}
}\ 
& 2\text{Re}\{\mathbf{b}_{5,g,j,k,n}^H \mathbf{\bar{B} }_{g,j,k,n}^H \mathbf{x} \}
\\
\textrm{s.t.}\ &
 \mathbf{x}^H\mathbf{x} \leq P.
\end{align}
\end{subequations}

Finally, 
we combine (\ref{Appendix_A_C3_6})$-$(\ref{Appendix_A_C3_7}) to derive the lower bound of
$\alpha_{g,j,n}$ is formulated in (\ref{MM_coefficient}).

Thus, the coefficients defined in (\ref{MM_coefficient}) have been proved.

%\vspace{-3.85em}

%\enlargethispage{-6.5cm}
\end{document}